\documentclass[12pt]{article}
\usepackage{graphicx}
\pdfoutput=1
\begin{document}

\title{Scaling forces to asteroid surfaces:  The role of cohesion}
\author{D.J. Scheeres, C.M. Hartzell, P. S\'anchez\\Department of Aerospace Engineering Sciences\\The University of Colorado\\Boulder, CO 80309-0429 USA\\scheeres@colorado.edu \\
M. Swift\\
University of Nottingham}
\date{}
\maketitle

\begin{abstract}
The scaling of physical forces to the extremely low ambient gravitational acceleration regimes found on the surfaces of small asteroids is performed.  Resulting from this, it is found that van der Waals cohesive forces between regolith grains on asteroid surfaces should be a dominant force and compete with particle weights and be greater, in general, than electrostatic and solar radiation pressure forces.  Based on this scaling, we interpret previous experiments performed on cohesive powders in the terrestrial environment as being relevant for the understanding of processes on asteroid surfaces.  The implications of these terrestrial experiments for interpreting observations of asteroid surfaces and macro-porosity are considered, and yield interpretations that differ from previously assumed processes for these environments.  Based on this understanding, we propose a new model for the end state of small, rapidly rotating asteroids which allows them to be comprised of relatively fine regolith grains held together by van der Waals cohesive forces.
\end{abstract}

\section{Introduction}

The progression of asteroid research, especially that focused on the smaller bodies of the NEA and Main Belt populations, has progressed from understanding their orbits, spins and spectral classes to more detailed mechanical studies of how these bodies evolve in response to forces and effects from their environment.  Along these lines there has been general confirmation that small NEAs are rubble piles above the 150 meter size scale, based both on spin rate statistics and on visual imagery from the Hayabusa mission to Itokawa.  However, the nature of these bodies at even smaller sizes are not well understood, with imagery from the Hayabusa mission suggesting that the core constituents of a rubble pile asteroid consists of boulders on the order of tens of meters and less \cite{fujiwara} while spin rate statistics imply that objects on the order of 100 meters or less can spin at rates much faster than seems feasible for a collection of self-gravitating meter-sized boulders \cite{HarrisPravec}.  Such extrapolations are based on simple scaling of physics from the Earth environment to that of the asteroid environment.  However, perhaps this is a process which must be performed more carefully.  In previous research, Holsapple \cite{holsappleA,holsappleB, holsapple_smallfast, holsapple_deform} has shown analytically that even small amounts of strength or cohesion in a rubble pile can render rapidly spinning small bodies stable against disruption.  In this paper we probe how the physics of interaction are expected to scale when one considers the forces between grains and boulders in the extremely low gravity environments found on asteroid surfaces and interiors.  

We note that asteroids are subject to a number of different physical effects which can shape their surfaces and sub-surfaces, including wide ranges in surface acceleration, small non-gravitational forces, and changing environments over time. Past studies have focused on a sub-set of physical forces, mainly gravitational, rotational (inertial) effects, friction forces, and constitutive laws \cite{holsappleA, holsappleB, holsapple_smallfast, holsapple_deform, richardson, Icarus_fission, sharma}.  Additional work has been performed on understanding the effect of solar radiation pressure \cite{burns} and electro-static forces on asteroid surfaces \cite{lee, colwell, hughes}, mostly motivated by dust levitation processes that have been identified on the lunar surface \cite{lunar_review}.  It is significant to note that the details of lunar dust levitation are not well understood.

The specific goal of this paper is to perform a survey of the known relevant forces that act on grains and particles, state their analytical form and relevant constants for the space environment, and consider how these forces scale relative to each other.  Resulting from this analysis we find that van der Waals cohesive forces should be a significant effect for the mechanics and evolution of asteroid surfaces and interiors.  Furthermore, we identify terrestrial analogs for performing scaled experimental studies of asteroid regolith and indicate how some past studies can be reinterpreted to shed light into phenomenon that occur on the surfaces of asteroids, the smallest aggregate bodies in the solar system.  

Taken together, our analysis suggests a model for the evolution of small asteroids that is consistent with previous research on the physical evolution and strength of these bodies.  In this model rubble pile asteroids shed components and boulders over time due to the YORP effect, losing their largest components at the fast phase of each YORP cycle and eventually reducing themselves to piles of relatively small regolith.  For sizes less than 100 meters it is possible for such a collection of bodies to be held together by cohesive forces at rotation periods much less than an hour.  
Finally, the implications of this work extends beyond asteroids, due to the fundamental physics and processes which we consider.  Specific applications of this work may be relevant for planetary rings and accretion processes in proto-planetary disks, although we do not directly discuss such connections.

The structure of the paper is as follows.  First, we review evidence for the granular structure of asteroids.  Then we perform an inventory of relevant forces that are at play in the asteroid environment and discuss appropriate values for the constants and parameters that control these results.  Following this, we perform direct comparisons between these forces and identify how their relative importance may scale with aggregate size and environment.  Then we perform a review of the experimental literature on cohesive powders and argue that these studies are of relevance for understanding fundamental physical processes that occur in asteroid regolith.  Finally, we discuss relevant observations of asteroids and their environment and the implications of our studies for the interpretation of asteroid surfaces, porosity and the population of small, rapidly spinning members of the asteroid population.  

\section{Evidence for the granular structure of asteroids}

Before we provide detailed descriptions of the relevant forces that act on particles and grains in the asteroid environment, we first review the evidence that has been drawn together recently which indicates that asteroids are dominated by granular structures, either globally or at least locally.

\subsection{Observations of asteroid populations}

For small asteroids, there are a few elements of statistical data that indicate the granular structure of these bodies.  
First is the size and spin distributions that have been tabulated over the years.  An essential reference is \cite{HarrisPravec}, which first pointed out the interesting relation between asteroid size and spin rate and provided the first population-wide evidence for asteroids being made of aggregates.  The naive implication of this is that larger asteroids are composed of distinct bodies resting on each other.  Thus, when these bodies reach sufficiently rapid rotation rates these components can enter orbit about each other and subsequently escape or form binaries \cite{Icarus_fission}.  The smaller components that escape, or conversely the larger asteroids that are eventually ``worn down'' by these repeated processes, then comprise a smaller population of what have been presumed to be monolithic bodies that can spin at elevated rates (although recent work has indicated that even small degrees of cohesion can stabilize these small bodies \cite{holsapple_smallfast}).  This has led to the development of the rubble pile model for asteroid morphologies with larger asteroids composed of aggregates of smaller bodies.  These smaller components are then available to comprise the population of fast spinning asteroids and range in size up to hundreds of meters.  

Second is the determination that asteroids can have high porosities in general.  The evidence for this has again been accumulated over many years, and has especially accelerated since the discovery of binary asteroids which allow the total mass, and hence density, to be estimated once a volume is estimated.  Porosity values have been correlated with asteroid spectral type \cite{B&C_AIII}, with typical porosities ranging from 30\% for S-type asteroids up to 50\% and higher for C-type asteroids.  Given good knowledge of the porosity of meteorite samples (on the order of 10\% in general) it is clear that asteroids must have significant macro-porosity in their mass distributions.  Existence of macro-porosity is consistent with a rubble pile model of asteroids, where there are components that have higher grain density resting on each other in such a way that significant open voids are present, leading to the observed macro-porosity.  This also motivates the application of granular mechanics theories to asteroids.

\subsection{Observations of specific asteroids}

Prior to the high resolution images of the surfaces of Eros and Itokawa, little was known about their small scale structure.  Eros shows fine-scale material with sizes much less than centimeters \cite{veverka_landing} with localized areas of very fine dust (presumed to be of order 50 microns) \cite{robinson}.  Itokawa shows a surface with minimum particle sizes at the scale of millimeters to centimeters \cite{yano} with evidence of migration of the finest gravels into the potential lows of that body \cite{miyamoto}.  Following these missions our conception of asteroid surfaces has changed significantly.  We now realize that the surfaces are dominated by loose regolith and that flow occurs across the surfaces of these bodies, causing finer materials to pool in the local or global geopotential lows of the body.  

In terms of geophysics, the important results from NEAR at Eros include the relatively high porosity (21-33\%) \cite{wilkison} along with a homogeneous gravity field, implying a uniform internal density \cite{miller,ask}.  For this body, which is large among NEA's, this implies a lack of large-void macro-porosity within its structure and instead a more finely distributed porosity throughout that body.  Observations of the surface of Eros have also enabled a deeper understanding of its constituents and internal structure.  By correlating degraded impact craters to physical distance from a recent, large crater on the surface of Eros, Thomas et al. \cite{Eros_crater_thomas} are able to show that seismic phenomenon from impacts are important for this body and cause migration of regolith over limited regions.  Support for this view comes from simulations carried out by Richardson et al. \cite{jim_richardson} which have attempted to determine a surface chronology for that body based on simple geophysics models.  Also, based on observations of lineaments across the surface of Eros, some authors have claimed that the body consists of a number of monolithic structures, perhaps fractured, resting on each other \cite{procktor, debra}.  Alternate views on interpreting surface lineaments have also been proposed, however, noting that they could arise from cohesion effects between surface particles \cite{asphaug_LPSC}.

The porosity of Itokawa was measured to be on the order of 40\%, and its surface and sub-surface seem to be clearly dominated by a wide range of aggregate sizes, ranging from boulders 10's of meters across down to sub-centimeter sized components.  The precision to which the asteroid was tracked precludes a detailed gravity field determination, as was done for Eros, thus we currently only have the total mass and shape of the body from which to infer mass distribution.  There is some tangential evidence for a non-homogenous mass distribution within the body, however, consistent with a shift in the center of mass towards the gravel-rich region of Itokawa, indicating either an accumulation of material there or a lower porosity \cite{itokawa_mdist}.  Another clear feature of the asteroid Itokawa is its bimodal distribution, allowing it to be interpreted as a contact binary structure.  The bulk shape of Itokawa can be decomposed into two components, both ellipsoidal in shape, resting on each other \cite{demura}.  
We also note that Itokawa has no apparent monolithic components on the scale of 100 meters, but instead appears to be rubble.  Another interesting result from the Hayabusa mission arises from the spacecraft's landing mechanics on the surface.  Analysis of the altimetry and Doppler tracking resulted in an estimated surface coefficient of restitution of 0.84 \cite{yano}, which is quite high for a material supposed to consist of unconsolidated gravels.  In \cite{miyamoto} observations of the Itokawa surface point to flow of finer regolith across the surface, pooling in the geopotential lows of that body.  Finally, size distributions of boulders on Itokawa show a dominance of scale at small sizes.  The number density of boulders is approximately $N \sim (r / 5)^3$ boulders per $m^2$, with $r$ specified in meters \cite{michikami}.  This leads to surface saturation at boulder sizes less than 12.5 cm.  

Although not a spacecraft rendezvous mission, significant results were also derived from the radar observations of the binary NEA 1999 KW4 \cite{KW4_ostro}.  Based on these observations, taken from the Arecibo radio antenna and the Goldstone Solar System Radar antenna, a detailed shape model for both components was created, the system mass determined, the relative densities of the bodies estimated, and the spin states of the asteroids estimated.  One item of significance is that the KW4 system is very similar to the majority of NEA binaries that have been observed \cite{pravecharris2006}.  A significant density disparity was found between the bodies, with the secondary having a mean density of 2.8 g/cm$^3$ and the primary a density of 2 g/cm$^3$.  Porosities of the primary body are estimated to be very high, with values up to 60\% being possible.  We also find that the primary is rotating at the surface disruption limit, near the rate where loose material would be lofted from its surface.  While the secondary is in a synchronous state, there is strong evidence that it is excited from this state, meaning that it is undergoing librations relative to its nominal rotation period which can cause relatively large variations in surface acceleration across its surface \cite{KW4}.  This environment was postulated to contribute to its low slopes and relatively high density.

Other, less direct, lines of evidence also point to the surfaces of asteroids as being dominated by loose materials.  First is the consistently low global slope distributions found over asteroid bodies at global scales.  Most of these results come from radar-derived shape models \cite{ostro_radar}, however they are also similar to the slope distributions found for Eros and Itokawa.  This is consistent with surfaces formed by loose granular material as granular dynamics predicts such limits on surface slope distributions.

Finally, a more recent result looked specifically for the signature of minimum particle sizes on asteroid surfaces by using polarimetry \cite{masiero}.  That paper observed a number of asteroids of similar type of different sizes and distances from the sun.  Applying the expected theory of dust levitation on asteroids \cite{lee,colwell} in conjunction with solar radiation pressure would predict that smaller grains should be absent from the surfaces of these bodies, and thus alter how light scatters from these surfaces.  Polarimetry observations of a number of asteroids did not yield any signature of minimum particle size differences on these bodies, however, and indicated a similar minimum size scale for surface particles independent of distance from the sun or size of the asteroid.  This is consistent with a lack of depletion of fines on surfaces, although there exist other explanations for this observation as well.  

\section{Physics of the Asteroid Environment}

We do not consider the strength of regolith grains and chondrules themselves, such as is implied in the strength-based models used in \cite{richardson_cohesion}, but only concern ourselves with the interaction between macroscopic grains and the environmental forces on these grains.  
Past studies have mostly focused on gravity and frictional forces, however it has also been speculated that, for particles at these size scales, electrostatic \cite{lee}, triboelectric \cite{marshall}, solar radiation pressure \cite{lpsc_SRP} and van der Waals' forces \cite{asphaug_LPSC} should also be included in that list.  Inclusion of these forces in studies of asteroid surfaces should have significant implications for the  mechanics of asteroid surfaces and for their simulation in terrestrial laboratories.  In addition to these forces, we will also include discussions of gravitational attraction between grains and on the pressures that grains will experience in the interiors of these bodies.
We assume spherical grains, which are generally used in granular mechanics studies due to the major simplifications this provides in analysis, and also due to demonstrated studies that this constitutes a reasonable model for the interactions of granular materials \cite{hermann, pingpong}.

\subsection{Ambient accelerations and comparisons}

\begin{figure}[h!]
\centering
\includegraphics[scale=0.12]{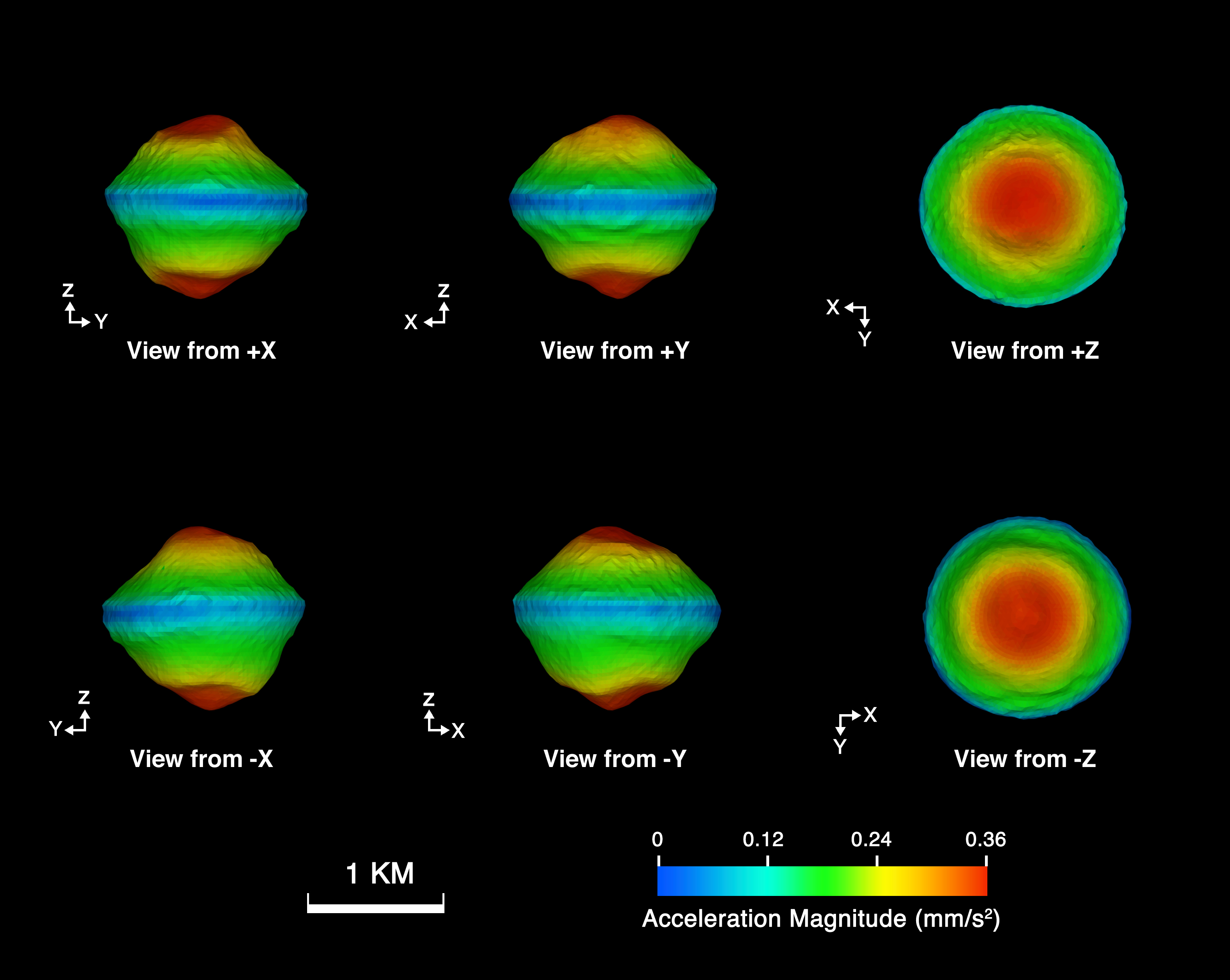}
\caption{Surface accelerations across the surface of the 1999 KW4 Primary.}
\label{fig:KW4}
\end{figure}

The most important defining quantity for this discussion is the ambient acceleration environment on the surface of a small body.  This consists of the gravitational attraction of the asteroid on a grain and the inertial effects that arise due to the rotation of the small body.  These effects generally act against each other, and thus reduce the ambient acceleration that grains on the surface of an asteroid feel.  The net effect of these competing effects can be substantial, as can be seen in Fig.\ \ref{fig:KW4} which shows the net gravitational accelerations across the surface of 1999 KW4 Alpha, the primary body of the binary asteroid 1999 KW4.  From this example we see that the surface acceleration can range over orders of magnitude, and thus the ambient environment for grains on the surface may have significant differences as one moves from polar to equatorial regions.

For understanding the relative effects of these forces on grains we will make comparisons between the grain's weight and the given force under consideration.  For an ambient gravitational acceleration of $g_A$ the ambient weight of a grain is defined as $W = m g_A$, where $m$ is the particle's mass ($4\pi/3 \rho_g r^3$ for a sphere), $\rho_g$ is the grain density (and is larger than the asteroid's bulk density), and $r$ is the grain radius.  In general we will assume $\rho_g = 3500$ kg/m$^3$ and will use MKS units throughout.

We find that the relevant forces acting on a grain are directly proportional to its radius elevated to some power.  Thus, a generic representation of a force acting on a grain can be given as $F = C r^n$, where $C$ is a constant and $n$ is an integer exponent in general.  A common representation of the strength of an external force used in granular mechanics is the bond number, which can be defined as the ratio of the force over the grain's weight:
\begin{eqnarray}
	B & = & \frac{F}{W} \\
	& = & \frac{3C}{4\pi\rho g_A} r^{n - 3}
\end{eqnarray}
In general $n \le 2$, meaning that our additional forces will usually dominate for smaller grain sizes.  We also note that the weak ambient accelerations will boost the bond numbers significantly, especially when we go beyond the milli-G regimes.  Using units of Earth gravity (1 G = 9.81 m/s$^2$) in order to describe the strength of ambient gravity fields, one milli-G equals $9.81\times10^{-3}$ m/s$^2$ and one micro-G equals $9.81\times10^{-6}$ m/s$^2$.  

\subsubsection{Gravitational and rotational accelerations}

Foremost for asteroid surfaces, and essentially controlling the environment by its strength or weakness, are gravitational and rotation induced centripetal accelerations acting on an asteroid and its surface.  If we model an asteroid as a sphere with a constant bulk density, the gravitational acceleration acting on a particle at the surface will be:
\begin{eqnarray}
	g & = & \frac{4\pi{\cal G}\rho}{3} R
\end{eqnarray}
where ${\cal G} = 6.672\times10^{-11}$ m$^3$/kg/s$^2$ is the gravitational constant, $\rho$ is the bulk density and $R$ is the radius of the body.  Introduction of non-spherical shapes will significantly vary the surface acceleration as a function of location on an asteroid, but will not alter its overall order of magnitude.  If we assume a bulk density of 2000 kg/m$^3$ (used for bulk density throughout this paper), we find that the surface gravity will be on the order of $5.6\times10^{-7} R$ m/s$^2$, or $\sim 5.6\times10^{-8} R$ G's.  Thus, a 1000 meter radius asteroid will have surface gravitational acceleration on the order of 50 micro-G's, scaling linearly with the radius for other sizes.

\begin{figure}[h!]
\centering
\includegraphics[scale=0.75]{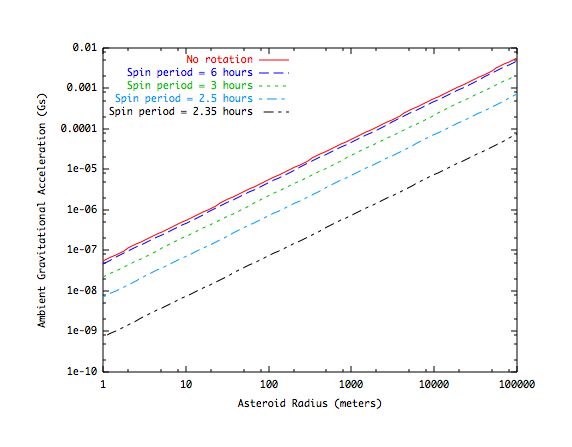}
\caption{Surface ambient gravity as a function of asteroid size and spin period.  Computed for a spherical asteroid with a bulk density of 2 g/cm$^3$.}
\label{fig:gravity}
\end{figure}

\begin{figure}[h!]
\centering
\includegraphics[scale=0.75]{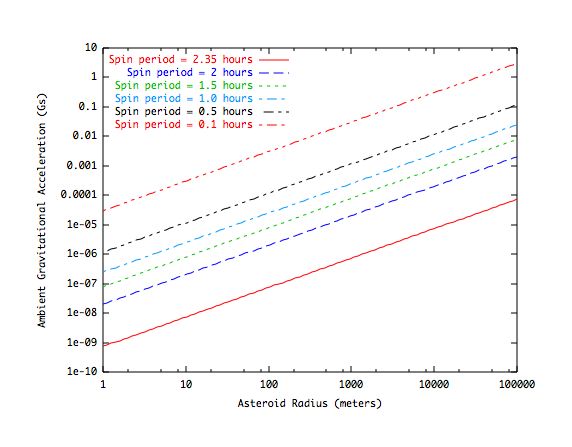}
\caption{Surface positive ambient gravity as a function of asteroid size and spin period.  Computed for a spherical asteroid with a bulk density of 2 g/cm$^3$.}
\label{fig:antigravity}
\end{figure}

Rotation also plays a significant role on the acceleration that a surface particle will experience.  Assume the asteroid is uniformly rotating about its maximum moment of inertia at an angular rate $\omega$.  Then at a latitude of $\delta$ (as measured from the plane perpendicular to the angular velocity vector), the net acceleration it experiences perpendicular to the rotation axis is $\omega^2 \cos\delta R$.  The acceleration it experiences normal to its surface due to rotation (assuming the asteroid is a sphere) is $\omega^2 \cos^2\delta R$.  Adding the gravity and inertial forces vectorially yields the net acceleration normal to the body surface:
\begin{eqnarray}
	g_A & = & \left( \omega^2 \cos^2\delta - \frac{4\pi{\cal G}\rho}{3} \right) R
\end{eqnarray}
with the largest accelerations occuring at $\delta = 0$.
We note that the centripetal acceleration acts against the gravitational acceleration, and that if the body spins at a sufficiently rapid rate particles on the surface can experience a net outwards acceleration, which is independent of the asteroid size and only dependent on its density.  For our chosen bulk density this rotation rate corresponds to a rotation period of $\sim$ 2.3 hours.  We note that an excess of asteroids have been discovered which are spinning at or close to this rate, and that those which spin faster tend to be smaller members of the population, with sizes less than 100 meters \cite{HarrisPravec}.  In Fig.\ \ref{fig:gravity} we show the relation between asteroid radius, spin period and ambient gravity at the equator for asteroids spinning less than their critical rotation period.  In Fig.\ \ref{fig:antigravity} we show the amount of ``cohesive acceleration'' necessary to keep a grain on the surface of an asteroid spinning beyond its critical rotation period.  We note that for asteroids of size 100 meters or less the radial outward accelerations are still rather modest, milli-Gs necessary for a 100 meter asteroid rotating with a 6 minute period or a 10 meter asteroid rotating with a period on the order of tens of seconds.

Incorporating these gravity and rotation effects for distended bodies yields significant variations over an object's surface.  For example, the total accelerations acting normal to the surface of Eros range from 0.2 to 0.6 milli-G's, on Itokawa these range from 6 to 9 micro-G's, and on the primary of the binary asteroid 1999 KW4 these accelerations range from 30 micro-G's to near zero (Fig.\ \ref{fig:KW4}).  These extremely low values of surface gravity set the stage for the other non-gravitational forces that can influence regolith on the surface.

\subsubsection{Coulomb Friction}

Intimately  linked with a particle's weight is the Coulomb frictional force.  The Coulomb force is proportional to the normal force between two grains and equals 
\begin{eqnarray}
	F_{F} & = & \mu N
\end{eqnarray}
where $\mu$ is the coefficient of friction and $N$ is the normal force.  For a particle resting on a surface and subject to no other forces, $N = W$. The physical nature of Coulomb friction arises from the mechanical interplay between particle surfaces and can have a component due to cohesion forces.  We discuss these combined effects later in the paper.  Coulomb friction plays a dominant role in describing the qualitative nature of surfaces, as this directly specifies the slope that a particle can maintain relative to  the body surface before sliding occurs.  This is the only one of the forces we consider that scales directly with ambient weight, with the coefficient of friction serving as the bond number for this effect, and implies that gravity and friction should be independent of size.  This particular result is sometimes invoked to claim that asteroid morphologies should scale independent of size, however our investigation of non-gravitational forces implies that this is not true.

\subsubsection{Interior Pressures}

Another important aspect for small bodies are their interior pressures.  Ignoring the rotation of the body, we can easily integrate across a spherical asteroid, assuming a constant bulk density of $\rho$, to find the pressures at a normalized distance ${\cal R}$ from the center (${\cal R}=1$ at the surface and 0 at the center):
\begin{eqnarray}
	P({\cal R}) & = & \frac{2\pi}{3}{\cal G} \rho^2 R^2 (1 - {\cal R}^2)
\end{eqnarray}
For the parameters assumed in this paper the pressure is
\begin{eqnarray}
	P({\cal R}) & = & 5.6\times10^{-4} R^2 (1-{\cal R}^2)
\end{eqnarray}
with units of Pascals.  Thus the pressures at the core of asteroids due to gravitational forces do not reach the kPa levels until we reach asteroids of radius 1300 m  and larger.  


\subsubsection{Self-Gravity}

When scaling forces down to the low levels we are considering, we should also consider the self gravitational force between two particles themselves.  Denote the two particles by their radius, $r_1$ and $r_2$ and assume they have a common grain density $\rho_g$ and are in contact.  Then the gravitational force between these two particles is
\begin{eqnarray}
	F_{self} & = & {\cal G} \left(\frac{4\pi\rho_g}{3}\right)^2 \frac{(r_1 r_2)^3}{(r_1 + r_2)^2}
\end{eqnarray}
For our assumed grain density value and equal sized particles we find the force between two particles to equal
\begin{eqnarray}
	F_{self} & = & 3.6\times10^{-3} r^4  \mbox{ [N]}
\end{eqnarray}
The bond number, defined for a particle of radius $r$ with grain density, is equal to
\begin{eqnarray}
	B_{self} & = & {\cal G} \frac{4\pi\rho_g}{3 g_A} r \\
	& \sim & 1\times10^{-6} \frac{r}{g_A}
\end{eqnarray}
We note that for micro-G environments ($g_A \sim 1\times10^{-5}$ m/s$^2$) boulders of 10 meter radius will have a unity bond number relative to self-attraction, increasing linearly with size.  Due to this scaling, we find that gravitational attraction between grains are close to the regime we are interested in, but can be neglected in general as we focus more on centimeter to decimeter sized particles.  However, we note that this local attraction effect could have significance for the interaction of larger collections of boulders and may imply that local interactions can be as important as the ambient field within which these boulders lie.

\subsubsection{Electrostatic forces}

Electrostatic forces have been hypothesized to play an important role on the surfaces of asteroids, and have been specifically invoked as one means by which small dust grains can be transported across a body's surface \cite{lee, robinson, colwell}.  These theories have been motivated by Apollo-era observations of dust levitation at the terminator regions of the moon \cite{lunar_review} and by the discovery of ponds on Eros \cite{robinson}.  Whether or not dust levitation occurs on asteroids is still an open question, although it is undoubtable that surface grains on these bodies are subject to electrostatic forces.  In the following we sketch out the main components of these electrostatic forces, including how they scale with particle size.  We only provide a limited discussion of the charges that particles can obtain, as this is still an active area of research and is not fully understood.

The electrostatic force felt by a surface particle is tied to its location. The charge accumulated at some point on the surface of an asteroid is due to an equilibrium reached between the current of electrons leaving the surface due to photoemission and the current of electrons impacting the surface from the solar wind. Both of these currents vary with location on the surface of the asteroid and with time as the asteroid rotates.  Photoemission is dependent on the solar incidence angle and solar wind interaction with the surface is dependent on a variety of plasma-related phenomena that vary with solar longitude. The resulting charge on the surface of the asteroid then influences the charging of the particle in question and influences the plasma environment (photoelectron and plasma sheaths) that will be experienced by the particle if it is lofted above the asteroid's surface. Thus, the first step in determining the electrostatic force experienced by a particle on an asteroid's surface is to determine the surface potential of the asteroid at that location. Following the procedure outlined in \cite{colwell} we find a surface potential for asteroids at the sub-solar point, $\phi_s$, equal to 4.4 V, holding relatively constant over a range of solar distances.
The surface potential of the asteroid can be directly related to the electric field \cite{colwell} as
\begin{eqnarray}
\label{eq:E0}
E=\frac{2 \sqrt{2} \phi_s}{\lambda_{D0}}
\end{eqnarray}
where $\lambda_{D0} \sim 1.4$ meters is the Debye Length of the photoelectron sheath.  The resulting electric field strength is $\sim9$ Volts/m, in agreement with both Lee and Colwell \cite{lee,colwell}.

To compute the force acting on a particle, it is necessary to specify the initial charge on the particle, however, there are significant uncertainties as to the exact charging mechanisms of particles in the space environment.
Given the charge, the electrostatic force acting on a particle is given by:
\begin{equation}
{F_{es}}=Q E
\label{eq:aes}
\end{equation}
where $Q$ is the total charge on the particle.
Should the particle have enough charge its electrostatic repulsion may cause it to levitate, or if it is lofted due to some other event, it will experience electrostatic forces throughout its trajectory near the asteroid's surface due to the charging of the particle and the surface.  We do not delve into these dynamics (c.f. \cite{colwell}), but instead focus on its environment on an asteroid's surface.

The charge on a particle is directly related to its potential and its radius as
\begin{eqnarray}
	Q & = & \frac{\phi_p r}{k_C}
\end{eqnarray}
where $\phi_p$ is the potential of the particle, $r$ is the particle radius and $k_C$ is the Coulomb constant.  To develop an estimate of the charging that a particle feels, we apply Gauss' Law to the asteroid surface.  This states that the total charge is proportional to the area that a given electric field acts over.  Specifically we use
\begin{eqnarray}
	Q & = & \epsilon_o E A
\end{eqnarray}
where $\epsilon_o$ is the permittivity, $E$ is the electric field and $A$ is the area in question.  Thus, as we consider smaller particles on the surface, with smaller areas, we expect the total charge of these particles to decrease.  Two implications can be found, for the potential of a particle and for the total force acting on it.  As the area of the particle varies as $r^2$, solving for the particle potential yields
\begin{eqnarray}
	\phi_p & \sim & \epsilon_o k_C E r
\end{eqnarray}
implying that the particle potential scales linearly with size.  Substituting the charge from Gauss' Law into the force equation provides
\begin{eqnarray}
	F_{es} & = & \epsilon_o E^2 A
\end{eqnarray}
Substituting the area of a sphere, $4\pi r^2$, we find the predicted force acting on a particle due to photoemission alone when directly illuminated to be
\begin{eqnarray}
	F_{es} & \sim & 4\pi \epsilon_o E^2 r^2
\end{eqnarray}
Given the permittivity constant in vacuum and the computed surface electric field we find the force acting on a particle of size $r$ to be
\begin{eqnarray}
	F_{es} & \sim & 9\times10^{-9} r^2
\end{eqnarray}
and the related bond number to be
\begin{eqnarray}
	B_{es} & \sim & 6\times10^{-13} \frac{1}{g_A r}
\end{eqnarray}
Thus for a micro-G environment we find a unity Bond number for particles of nanometer size and conclude that electrostatics due to photoemission alone is negligible.

The same situation may not exist in the terminator regions of the asteroid surface, however.  
Hypothesized mechanisms for spontaneous dust levitation have relied on enhancements to the nominal charging environment to generate sufficient charge or electric field to move particles off of an asteroid's surface \cite{lee, colwell}.  Explanations for dust levitation on the moon have relied on effects active at the terminator to focus the electric fields and raise them to sufficiently high values to overcome gravitational attraction \cite{criswell}.  In scaling the resulting electric fields to asteroid terminators, Lee estimates that  large electric fields on the order of $10^5$ V/m can occur, substantially enhancing the relevance of electrostatics.  Similarly, triboelectric charging of particles may be able to generate large voltages of comparable size.  Such charging conditions have not been verified in the space environment, although they are sufficient to increase the relevance of electrostatic forces.  We borrow the results from Lee to generate an estimate of possible electrostatic forces on asteroids in the vicinity of their terminators.  Using these stronger electric fields in our above analysis provides forces on the order of
\begin{eqnarray}
	F_{es} & \sim & 0.1 r^2
\end{eqnarray}
for particles resting on the surface.  Although unverified, we will use this force as representative of the maximum strength of electrostatic forces acting on a particle on an asteroid surface.  The bond number for these larger forces are
\begin{eqnarray}
	B_{es} & \sim & 7\times10^{-6} \frac{1}{g_A r}
\end{eqnarray}
Thus, in the enhanced regimes that have been hypothesized to exist at terminators, in a micro-G environment particles of radius 0.7 meters have unity bond numbers.

\subsubsection{Solar radiation pressure forces}

Whenever a particle is subjected to full illumination by the sun, photons are reflected, absorbed and re-emitted from grains.  This can occur when the particles lie on the surface, but become more significant if the grain is lofted from the surface of the asteroid.  The photon flux provides a pressure that acts on the grain which is easily converted to a force.  The physics of dust grain-photon interactions are studied in Burns et al. \cite{burns}, where relativistic and scattering effects are considered in detail.  For our current study we focus mainly on grains on the order of microns or larger, where geometric optics derived results describe the force acting on such grains.  For grain sizes less than 0.5 microns, the interactions of dust particles with solar photons becomes more complex due to the maximum flux of the sun occurring at wavelengths of commensurate size to the particles themselves, reducing the efficiency of momentum transfer.  

For this simple geometrical optics model, we find the force acting on a particle to be
\begin{eqnarray}
	F_{srp} & = & \frac{{\cal G}_{SRP}(1+\sigma)}{d^2} A
\end{eqnarray}
where ${\cal G}_{SRP} \sim 1\times10^{17}$ kg m/s$^2$, $A$ is the illuminated particle area and $d$ is the distance to the sun.  We choose the term $\sigma$ to generally represent the effect of reflection, reemission or loss of coupling.  Specifically, $\sigma = 1$ for a fully reflective body, equals $2/3$ for a body that reflects diffusively, is zero for an absorbing body that uniformly radiates and is negative (but greater than -1) for small grains that decouple from the maximum solar radiation flux at visible wavelengths \cite{burns}.  The force that a particle feels varies as $r^2$, for an asteroid at 1 AU from the sun the specific values are 
\begin{eqnarray}
	F_{srp} & = & 1.4(1+\sigma) \times10^{-5} r^2
\end{eqnarray}
We note that this force dominates over the electrostatic force we find using a simple balance of photoemission currents, but is much smaller than the hypothesized forces due to enhanced electric fields at an asteroid terminator.  They both share the $r^2$ dependence, however.  The Bond number for this force is computed to be
\begin{eqnarray}
	B_{srp} & = & 1\times10^{-9} \frac{1+\sigma}{g_A r}
\end{eqnarray}
Thus, for a micro-G environment the Bond number is unity for grain radii on the order of 100 microns.

The dynamics of particles in orbit about an asteroid are subject to major perturbations from SRP, and for many situations the SRP forces can exceed gravitational attraction and directly strip a particle out of orbit.  These dynamics have been studied extensively in the past, both at the mathematical and physical level \cite{mignard, richter, D, dankowicz}.  The relevance of these forces when on the surface of a body have not been considered in as much detail, but could be a significant contributor to levitation conditions, both hindering and helping levitation depending on the geometry of illumination.  

\subsubsection{Surface Contact Cohesive Forces}

Finally we consider the physics of grains in contact with each other and exerting a cohesive force on each other due to the van der Waals forces between individual molecules within each grain.  The nature and characterization of these forces has been investigated extensively in the past, both experimentally and theoretically \cite{johnson, heim, castellanos}.  There is now an agreed upon, and relatively simple, theory that describes the strength and functional form of such contact cohesion forces \cite{rognon, castellanos}.  Despite this, a detailed discussion of such forces for the asteroid surface has not been given as of yet, although the implications of these forces for lunar regolith cohesion has been investigated \cite{perko}.  We take the lunar study as a starting point for applying the theory to asteroid surfaces.

The mathematical model of the van der Waals force which we adopt is rather simple \cite{castellanos, perko, rognon}, and for the attraction between two spheres of radius $r_1$ and $r_2$ is computed as:
\begin{eqnarray}
	F_c & = & \frac{A}{48 (t+d)^2} \frac{r_1 r_2}{r_1 + r_2} 
\end{eqnarray}
where $A$ is the Hamaker constant and is defined for contact between different surfaces in units of work (Joules), $t$ is the minimum inter-particle distance between surfaces and is non-zero in general due to the adsorption of molecules on the surfaces of these materials, $d$ is the distance between particle surfaces, and $r = r_1 r_2 / (r_1 + r_2)$ is defined as the reduced radius of the system.  The details of these interactions have been extensively tested in the laboratory across different size scales \cite{johnson, heim}.  It is also important to note that the attractive force is relatively constant, independent of particle deformation \cite{derjaguin, maugis}, meaning that this simple form of the cohesive forces can be used as a general model for particles in contact with each other without having to explicitly solve for particle deformation.  The Hamaker coefficient $A$ tends to be so small that the cohesion force effectively drops to zero for values of distance $d$ between the surfaces of the particles on the order of particle radii, thus we generally suppress this distance term $d$ in the following and only consider the force to be active when the bodies are in contact (see \cite{castellanos} for more details).

In the space environment the minimum distance between the materials, $t$, can be much closer than possible on Earth where atmospheric gases, water vapor, and relatively low temperatures allow for significant contamination of surfaces.  In the extreme environment of space, surfaces are much ``cleaner'' due to the lack of adsorbed molecules on the surfaces of materials \cite{perko}, allowing for closer effective distances between surfaces.  Perko defines the cleanliness ratio as the diameter of an oxygen molecule divided by the thickness of the adsorbed gas on the surface of a sample.  In terrestrial environments this cleanliness ratio can be small, due to the large amount of gas and water vapor that deposits itself on all free surfaces.  In low pressure or high temperature conditions, however, as are found on the moon and on asteroid surfaces, cleanliness ratios can approach values of unity, meaning that particle surfaces can come in extremely close contact with each other, essentially separated by the diameter of their constituent mineral molecules.  In these situations the strength of van der Waals forces can become stronger than are experienced between similar particles on Earth.  For lunar soils at high temperatures Perko finds that cleanliness ratios approach unity, meaning that the distance $t \rightarrow 1.32\times10^{-10}$ meters.  Following \cite{perko} we define the surface cleanliness as $S \sim \Omega / t$, where $\Omega \sim 1.5\times10^{-10}$ meters and $t$ is the minimum separation possible between two particles in contact.  A clean surface, typical of lunar regolith on the sun-side, can have $S\rightarrow 1$, while in terrestrial settings in the presence of atmosphere and water vapor we find $S \sim 0.1$ \cite{perko}.  Modifying the cohesion force incorporating the surface cleanliness ratio and setting $d \sim 0$ we find
\begin{eqnarray}
	F_c & = & \frac{A S^2}{48 \Omega^2} r 
\end{eqnarray}

In the following we use the appropriate constants for lunar regolith, a Hamaker constant of $4.3\times10^{-20}$ Joules and an inter-particle distance of $1.5\times10^{-10}$ meters \cite{perko}.  This is conservative in general, meaning that these will provide under-estimates of the van der Waals force for particles on asteroids or in micro-gravity, as they are computed for the surface of the moon where there is still some remnant atmosphere contributing to surface contamination and hence larger values of $t$.  These combine to yield an equation for the van der Waals force at zero distance ($d=0$):
\begin{eqnarray}
	F_{c} & = & 3.6\times10^{-2} S^2 r
\end{eqnarray}
This formula reconstructs the measured cohesive forces determined by Perko \cite{perko}.  The Bond number for this force equals
\begin{eqnarray}
	B_c & = & 2.5\times10^{-6} \frac{S^2}{g_A r^2}
\end{eqnarray}
For a micro-G environment we find unity Bond numbers at particle radii on order of 0.5 m, and thus we note that this force is significant.  

A final consideration is the net effect of heterogeneous size distributions within cohesive aggregates, and the closely related effect of surface asperities or irregularities of individual grains.  All experimental tests generally deal with size distributions of irregularly shaped grains, and thus the results we find from these tests should be informative for realistic distributions found at asteroids.  This being said, the potential size scales over which cohesive forces are relevant may be much wider at asteroids, and thus could result in effects not seen in Earth laboratories.  Castellanos studies the effects of surface asperities and the inclusion of relatively small particles within printing toners and analytically characterizes their effects on cohesive forces \cite{castellanos}.  Summarizing the detailed results of that study, we find that the net effect of surface asperities on a particle will change the cohesive force scaling from the particle radius to the asperity radius, $r_a$, which can easily be up to an order of magnitude smaller than the particle radius.  Similarly, a particle that is covered with many smaller particles, with radius denoted as $r_a$ again,  will interact with neighboring particles (with a similar coating) with a cohesive force proportional to $r_a$.  Thus, consider a ``clean'' particle of radius $r$ covered by asperities or smaller particles of radius $r_a$.  An approximate equation for the cohesive force can then be represented as
\begin{eqnarray}
	F_{ca} & \sim & F_{c} \frac{r_a}{r}
\end{eqnarray}
where $r_a < r$ in general.  We can use Perko's cleanliness ratio as a qualitative parameter to account for this effect by letting $S \sim \sqrt{r_a / r}$.  Thus a cleanliness ratio of 0.1 can be related to a grain being coated by particles that are one-hundredth its size.  Alternately, a grain with surface asperities about one-tenth of the grain size will have a cleanliness ratio of about 0.3.  The details and physics of these corrections are more involved than the simple scaling we use here, although they follow these general trends \cite{castellanos}.  For a fixed macroscopic grain size, as the asperities or smaller particles shrink in size, the reduction in cohesion force does not go to zero, but becomes limited due to the disparity in size between the macroscopic grain and the smaller features, as the size of these features begin to become small relative to the local surface curvature of the grain.  

\subsection{Scaling particle forces to the asteroid environment}

Having defined the relevant known physical forces that can act on surface particles, we can make direct comparisons of these forces to ascertain which should dominate and in what region.  

\subsubsection{Direct Comparisons with Gravitational Forces}

\begin{figure}[h!]
\centering
\includegraphics[scale=0.75]{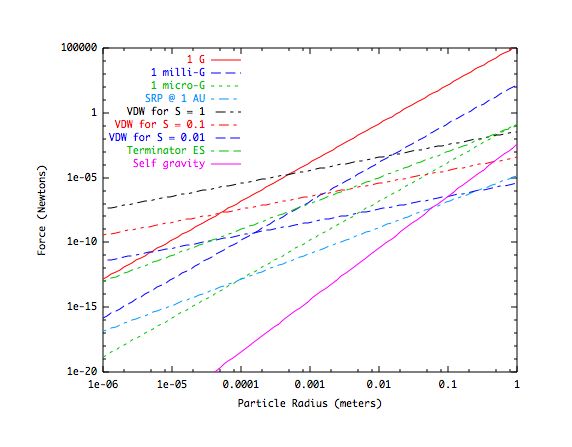}
\caption{Comparison of forces for surface particles of different radii.}
\label{fig:force}
\end{figure}

\begin{figure}[h!]
\centering
\includegraphics[scale=0.75]{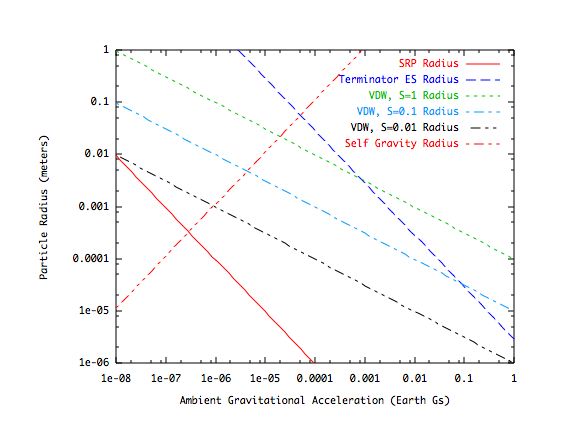}
\caption{Radii of surface particles for weight equal to force as a function of ambient $G$.}
\label{fig:radius}
\end{figure}

First we note some simple scaling laws that are at play for the relevance of non-gravitational forces to gravitational forces.  As is well known, gravitational (and rotational) accelerations are constant independent of particle size, but as forces will vary with the mass of the object, i.e., as $r^3$ with the radius of the particle.  We note that the self-gravity, solar radiation, electrostatic and cohesion forces all vary with a different power of particle radius.  Self-gravity varies as $r^4$, solar radiation pressure and electrostatic as $r^2$, and cohesion as $r$.  Thus when comparing these forces with ambient gravitational force, we find that the forces take on different levels of significance for different particle sizes.  To characterize these relationships we compute the forces as a function of particle size in Fig.\ \ref{fig:force} and compute the particle size at which ambient weight equals force as a function of ambient acceleration in Fig.\ \ref{fig:radius}.  We note that some of these forces are attractive, some repulsive, and some depend on the relative geometry of the grains.  Thus we only compare the magnitudes.

\subsubsection{Self-Gravity and Cohesion}

We first consider a direct comparison between the self-gravitational attraction of two spheres in contact as compared with their predicted cohesive attraction.  Solving for the radius where the predicted cohesion equals self attraction between two particles we find $r = 10^{1/3} S^{2/3}$.  For clean surfaces this radius is approximately 2 meters while for cleanliness ratios of 0.1 and 0.01 it reduces to 0.5 and 0.1 meters, respectively.  In the context of our ambient gravitational environments, we see that self-gravity falls outside of the forces of most interest to us, however it is surprisingly close to our regime.  Our detailed discussions will be focused on particles with sizes on the order of tens of centimeters and smaller later in this paper, and thus we note that self-gravitation between particles is not quite relevant for these sizes.  For meter-class bodies, however, we note that cohesiveness and gravity are of the same order, which could be an important consideration for the global mechanics of rubble pile asteroids and a topic for future research.  

\subsubsection{Solar Radiation Pressure and Cohesion}

For solar radiation pressure, we note from \cite{burns} that particles much less than one micron are in general invisible to radiation pressure.  Thus, we see that surface particles are not significantly perturbed by SRP until we get below the milli-G level.  In terms of gravitation, this occurs for micron-sized particles at an asteroid radius of 18 kilometers and increases to centimeter-sized particles for rapidly rotating asteroids at tens to hundreds of meters.  We also note that the plots predict that cohesive forces dominate over solar radiation pressure across all of these particle sizes on the surfaces of asteroids.  Solving for the radius at which SRP and cohesion are equal we find a surprisingly large value of 100 meters in radius, although we note that the application of cohesive forces for such a large object may not be realistic.  Still, this simple scaling indicates that SRP may not be a relevant force on the surface of asteroids, even though it can play a dominant role once a particle is lofted above the surface and the cohesive force removed.

\subsubsection{Electrostatics and Cohesion}

A direct comparison of forces between electrostatics and cohesion, assuming the terminator charging electric field strengths of $10^5$ V/m, yields equality for particle radii of order $0.3 S^2$ meters, where smaller sized particles will be dominated by cohesion.  The uncertainties in the strength of terminator electric fields and the charging of surface particles provides a large range of uncertainties on this estimate.  However, increases in electric field strength by a factor of 6, well within the range of uncertainties discussed in \cite{lee}, will create equal forces for particle sizes on the order of 1 cm, implying that the strong fields at terminators may be able to break cohesive forces and directly levitate larger grains.  This comparison does point out challenges for directly levitating small grains from the surface of an asteroid in the absence of some other mechanism, however.  It is also important to note that these levitation conditions require specific shadowing environments and other local conditions, and thus are not ubiquitous globally.  

\subsubsection{Cohesion and Ambient Gravity}

In the milli to micro-G range, we find that cohesive forces become important for particles of radius 1 cm up to 1 meter in size and smaller.  Again, simple scaling to these sizes is more complicated than these comparisons, yet this indicates that regolith containing grains of millimeter to decimeter sizes may undergo significantly different geophysical processes than similar sized particles will in the terrestrial environment.  In fact, that asteroid regolith may be better described by cohesive powders (for a familiar analogy, consider the mechanical properties of bread flour) than by traditional analyses of landslides.  Thus, after these comparisons we conclude that a reasonable analog for asteroid regolith are cohesive powders, which have been studied extensively in the 1-G environment for practical applications on Earth.
In Table \ref{tab:1} we list the size of grains for unity bond numbers as a function of different ambient accelerations, and note the bodies at which these ambient accelerations are found.

\begin{table}[h!]
\centering
\begin{tabular}{|| c || c || c ||}
\hline
Gravity ($G$s) & Grain Radius (meters) & Analog body \\
\hline
\hline
1 & 		$6.5\times10^{-4}$ & Earth \\
0.1 &	$2\times10^{-3}$ & Moon \\
 0.01 & 	$6.5\times10^{-3}$ & Vesta (180 km) \\
 0.001 (milli-G)& 	$2\times10^{-2}$ & Eros (18 km) \\
 0.0001 & 	$6.5\times10^{-2}$ & Toutatis (1.8 km) \\
 0.00001 & 	$2\times10^{-1}$ & Itokawa (0.18 km) \\
 0.000001 (micro-G)& 	$6.5\times10^{-1}$ &  (0.018 km) \\
 0.0000001 & 	$2\times10^{0}$ & KW4 Equator \\
\hline
\hline
\end{tabular}
\caption{Radius at which ambient weight and cohesion forces are equal (assuming lunar regolith properties), along with nominal parent body sizes.}
\label{tab:1}
\end{table}

\section{Experimental and Theoretical Results and their Implications for Asteroids}

One main impetus behind this article is to develop the basic scaling relations between asteroid regolith and cohesion effects in order to motivate terrestrial testing of regolith properties through the use of appropriate materials.  Specifically, previous research has tacitly used regolith models chosen to emulate gravels and other coarse material, based on the visual interpretation of asteroid surface morphology \cite{miyamoto, procktor, debra}.  However, the proper terrestrial analogue in terms of local properties may be much more similar to cohesive powders, as has been surmised by Asphaug \cite{asphaug_LPSC}.  With this change in perspective, we can access previous literature and testing for cohesive powders and reinterpret them as indicative of asteroid regolith properties, especially for small bodies that have regolith in the milli to micro-G regime.  This being said, the literature on the granular mechanics properties of cohesive powders is relatively limited, especially for those studies of relevance to asteroid regolith.  However we find that studies of the granular mechanics behavior of cohesive powders exhibit a variety of outcomes that mimic observed asteroid behaviors.  

Cohesiveness can be imparted to granules in two basic ways, first is to add fluid to an existing granular material.  Second is to grind granular materials to small enough size for van der Waals forces to become effective.  It is only the latter that are relevant for understanding asteroid surfaces.  Indeed the response of materials made cohesive in these two different ways have been observed to have significantly different mechanical and dynamical properties \cite{alexander}.  In the following we cite some recent research in the field of cohesive mechanics and note analogs for observations in the field of asteroid mechanics.  We first provide brief summaries of the recent research that we draw from.

{Perko} \cite{perko} details a theoretical and experimental analysis that characterizes the cohesive properties of lunar regolith.  The important results from that paper, some of which have already been used, are the concept of surface cleanliness and its relation to cohesion, a soil mechanics analysis of lunar regolith accounting for van der Waals cohesive forces, and fundamental data on the cohesive properties of lunar regolith which we have adopted to serve as a model for asteroid regolith.

{Alexander}  \cite{alexander} details comparisons between cohesive powder flows and numerical simulations, and measures several important results for the avalanching behavior of cohesive powders.  Most relevant for our work is the measured onset of bulk cohesive effects and the measured dilation of avalanching flows.

{Rognon} \cite{rognon} provides the results of a number of detailed numerical simulations that describe the dynamics of flowing cohesive grains.  This study independently varies the cohesion between grains and the inertia number (i.e., flow velocity) of granular materials.  As it is a set of numerical computations they are able to extract a wide range of relevant statistics that provide insight into cohesive powder flows.

{Meriaux} \cite{meriaux} describes experiments in which columns of cohesive material were formed and then caused to collapse suddenly (by removal of a supporting wall) or in a quasi-static fashion (by slowing moving a barrier wall).  The bond numbers of their granular materials are not given, but the cohesive nature of their powders were verified experimentally.  The main independent parameter of their experiments were the aspect ratio of their initial columns, defined as the height of the column divided by its one-dimensional length, with the third dimension (depth) being held fixed.  The observable outcome, besides observations on the granular material morphology, was the final height of the column and the final runout length of the column.  

{Vandewalle}  \cite{vandewalle} describes a series of experiments that investigated the compaction of material subjected to repeated taps.  While not exclusively focused on cohesive materials, there are a number of relevant observations for the compaction and flow of cohesive powders.

\subsection{Onset of Macroscopic Cohesive Effects}

\begin{figure}[h!]
\centering
\includegraphics[scale=0.75]{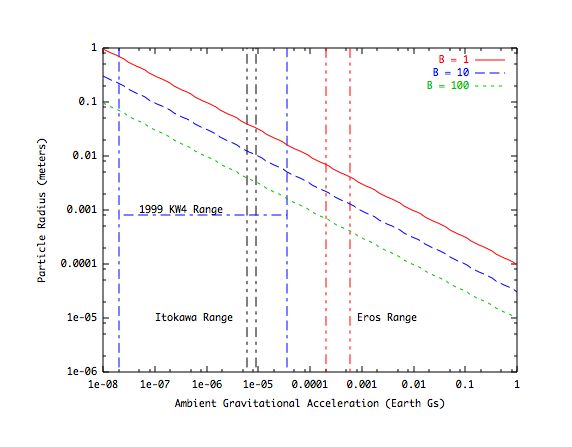}
\caption{Particle radii for different Bond numbers, assuming lunar regolith properties with surface cleanliness $S=1$.}
\label{fig:bond}
\end{figure}

As compared to the flow and mechanics of non-cohesive aggregates, several key issues arise when cohesive forces become relevant for the mechanical regimes of interest.  First, we note that the strength of cohesive forces are often parameterized by the bond numbers introduced earlier, where a bond number of 1 means that the force of cohesion equals weight.  For the simulation of global mechanical properties of cohesive powders we find that modelers often use bond numbers on the order of 10-30 or larger to observe macroscopic behavior \cite{alexander, rognon}.  At bond numbers of 100 and greater, experiments have shown that particles will preferentially stick to each other and form clumps of materials, which can then flow and act as larger particles.  Figure \ref{fig:bond} shows particle radius vs.\ ambient gravity for different cohesive bond numbers, with the surface gravity of Eros, Itokawa and 1999 KW4 Alpha indicated. 

We note that for the Itokawa environment bond numbers of 100 correspond to few millimeter-sized grains.  The highest resolution images of the Itokawa surface indicate grains of centimeter size, allowing a different interpretation of that surface as not of being composed of competent grains of this size but instead of smaller sized grains that are preferentially clumped at this size scale.  For Eros, this clumping behavior would dominate at 1 millimeter grains and smaller, which were well below the resolution limit of the highest resolution images taken by NEAR.  At the low end of the 1999 KW4 environment (along the equator) we find bond numbers of 100 at the several centimeter level.  It is not clear how such large particles would interact with each other at these low G levels, we do note that the strongest predicted electrostatic forces should begin to dominate at these size scales and that the presence of smaller and finer regolith could also influence the overall cohesive strength between such large grains (characterized by the surface cleanliness).  At this point, we are only able to point out the scaling regime where these materials fall, and must wait for high resolution images of these regions and mechanical tests of asteroid surfaces (presumably from spacecraft) in order to better understand how materials will interact with each other at these extremely low ambient gravity levels.

\subsection{Effects of Cohesion on Shear Strength}

Cohesion forces arising from van der Waals effects also modify the expected shear strength of asteroid regolith, and can create a size dependance on these effects.  
From a classical mechanics perspective, the effect of cohesion and porosity on a granular material's yield criterion is reviewed in \cite{schwedes}, where a three dimensional ``condition diagram'' is presented as a general approach to describing how a cohesive granular material will fail as a function of compressive stress, shear stress and porosity.  Analysis of the failure surfaces directly indicate how a body undergoing failure will often dilate, as will be discussed later.  Despite the existence of this general approach to describing the yield failure of a granular material, recent analyses have focused on more direct measures, such as the internal angle of friction, additional cohesive forces, and other bulk characterizations of material properties.

We directly discuss two different approaches to this topic.  First we consider Rognon's numerical investigations of the constitutive law as a function of bond number.  Then we discuss Perko's analysis of lunar regolith and his scaling of their properties to distributions.  

Rognon \cite{rognon} studies the constitutive relationship for cohesive granular flows numerically.  His full analysis considers the variation of the friction coefficient as a function of the inertia number of the flow, however in the presence of strong cohesion the dependence on the inertia number becomes subdued.  Thus we only consider his quasi-static expression for the shear stress, expressed in Perko's notation as:
\begin{eqnarray}
	\tau & = & \mu \sigma_n + c
\end{eqnarray}
where $\mu = \tan\phi$ is the friction coefficient, $\sigma_n$ is the normal stress and $c$ is the additional cohesive stress.  Rognon analyzes the relationship between the friction coefficient and bond number, finding a near linear growth in $\mu$ with bond number, starting at less than 0.5 at zero bond number and increasing to $\sim 1.5$ at a bond number of 80, in this way noting the ability of cohesive grains to sustain larger slope angles.  The additional cohesive stress $c$ is modeled as:
\begin{eqnarray}
	c & = & \beta \frac{F_{c}}{r}
\end{eqnarray}
where $\beta$ is numerically determined to equal 0.012 for flowing material.  Predictions from Coulomb theory are that $\beta \sim 0.2$ \cite{rognon}.  In Rognon's analysis this difference between numerically determined and predicted cohesive stress occurs due to the grains agglomerating into larger aggregates which are able to flow across each other more easily.  Given that Rognon's analysis is more relevant for flow of granular material on a surface, this self-organizing behavior may not be as relevant beneath the surface or for understanding the soil mechanics aspects of cohesive grains.

Perko \cite{perko} characterized the effect of cohesion forces as an addition to the existing bulk cohesion stress and friction angle of a given sample.  This formulation was chosen as it allows him to describe the variation in cohesive properties as a function of time (i.e., incident sunlight) on the lunar surface.  As such, he characterizes the shear strength as
\begin{eqnarray}
	\tau & = & c + c' + \sigma_n\left( \tan\phi + \tan\phi'\right)
\end{eqnarray}
where $c$ represents cohesion, $\phi$ is the friction angle, and $\sigma_n$ is the effective normal stress.  The primes denote additional cohesion and friction angle contributions due solely to van der Waals effects as the surface cleanliness is increased.  The normal stress is computed as
\begin{eqnarray}
	\sigma_n & = & \eta N \cos\alpha
\end{eqnarray}
where $N$ is the normal force, $\alpha$ represents the angle between the resultant of the normal force and the direction of $\sigma_n$ and can range up to 30$^\circ$, and $\eta$ is the number of particle contacts per unit area.  Perko relates $\eta$ and porosity by a simple scaling,
\begin{eqnarray}
	\eta & \sim & \frac{P \xi}{4 r^2}
\end{eqnarray}
where $P$ is a porosity factor (not porosity) varying from 0.6 for loose material to 4 for dense soils, $\xi$ is an angularity factor and ranges from 1 for spheres to 8 for rough particles, and $r$ is the radius of the particles under consideration.  
The additional contributions to cohesion and friction angle are:
\begin{eqnarray}
	c' & = & F_c \eta \\
	\tan\phi' & = & \frac{A}{48\pi\sigma_y t^3\cos\alpha}
\end{eqnarray}
where $\sigma_y$ is the contact yield stress, which we do not consider in detail.

Lunar regolith in the upper 15 cm has a cohesion of $c \sim 5$ kPa and $\phi \sim 41^{\circ}$ \cite{colwell},  although it is not clear what fraction of the cohesion value is due to van der Waals forces.  
For lunar regolith in the daytime, when the surfaces have a higher level of cleanliness, Perko estimates the additional cohesion to be 0.5 kPa and the additional friction angle to be $24^\circ$, computed for an average diameter of 70 microns, $P = 0.9$ and $\xi = 2$.  

Generalizing this result to arbitrary grain radii we find $\eta = 0.45 / r^2$ contacts per m$^2$.  Thus, for a sphere with surface area $4\pi r^2$ this predicts $\sim 6$ contacts per particle, independent of size.  The numerical factor in $\eta$, 0.45, can be compared with Rognon's $\beta$ and we see an order of magnitude difference in their estimated values.  Recall that Rognon's estimate is a numerical computation for dynamically flowing material and Perko's is an experimental measurement for soil, which could explain why the results are different.  However, such mismatches also indicate the uncertainties associated with this field.  Although an order of magnitude difference appears to be significant, as our current analysis is looking at ranges of particle size and ambient acceleration such a difference does not change our overall qualitative conclusions.

Applying the force constants for lunar regolith, the additional cohesion contribution due to van der Waals forces is estimated to be
\begin{eqnarray}
	c' & = & 1.6\times10^{-2} \frac{1}{r} \mbox{ [Pa]}
\end{eqnarray}   
Thus we find that the additional cohesive shear contribution is 1 Pa at 1.6 centimeter sizes and 1 kPa at 16 microns.  The additional friction angle, as stated in \cite{perko}, is independent of grain size and equal to $24^\circ$.  The normal stress is a function of $\eta$, grain mass and ambient gravity.  Combining these effects, and taking $\alpha = 0^\circ$, the additional frictional shear is estimated to be
\begin{eqnarray}
	\sigma_n \tan\phi' & = & 3\times10^{3} r g_A  \mbox{ [Pa]}
\end{eqnarray}
For a 1 meter particle in a 1-G field, the frictional shear is 30kPa, while for a 1 meter particle in a micro-gravity regime it reduces to 0.03 Pa and is vanishingly small for millimeter and smaller grains.  If, instead, we use the normal stresses found in the interior of a small body, the frictional stresses will be independent of grain mass and the additional frictional shear due to cohesion will be on the order of 
\begin{eqnarray}
	\tau' & = & 2.5\times10^{-4} R^2
\end{eqnarray} 
where $R$ is the asteroid radius in meters.  Thus, the additional strength due to cohesion can reach values of 1 kPa for asteroids of size 2 km and larger, independent of grain size.  Depending on the minimum grain size in the asteroid interior, we can find additional cohesive shear strength on the order of kPa.

For surface particles the main implications are that cohesive shear is enhanced for small grains while cohesive friction is enhanced for larger grains.  Thus, at larger sizes we expect an enhanced slope, increasing from $41^\circ$ to $52^\circ$ in the presence of cohesion.  Conversely, at the finer scales which are, as of yet, unexplored for asteroid surfaces, we would expect much stronger local topography, with an ability to create rough terrains due to enhanced cohesiveness.  We note that previous assertions of smooth regions on asteroid surfaces, in particular the ponds on Eros and the seas on Itokawa, have been based on observations at relatively coarse resolutions, reaching centimeters at best, and then only at low phase angles.  Sub-millimeter observations of these surfaces should reveal the small scale strength of local topography on an asteroid.  

Regarding the implications of global enhancements to shear, we refer to  Holsapple \cite{holsapple_smallfast} where he finds that additional shear strength on the order of a few to 10 kPa can keep a small body's shape stable against very rapid spins.  The connection between Holsapple's shear model and the current model should be explored and understood in the future, but is not addressed in detail here.  

\subsection{Flows of Cohesive Materials}

For flowing granular materials a key parameter is the ``Inertia Number'', which compares the relative importance of shear rate in a flow and pressure.  Following \cite{meriaux} we compute this as:
\begin{eqnarray}
	{\cal I} & = & \frac{U}{\sqrt{g_A H}} \frac{r}{L}
\end{eqnarray}
where $U$ is the speed of the flow, $g_A$ is the ambient acceleration, $H$ is the altitude/depth of the granular material, $r$ is the size of the grains, and $L$ is a characteristic length that the granular material is distributed over.  This is also interpreted as the ratio of the confinement pressure timescale over the shear deformation timescale.  It can be shown that a system of freely sliding particles down a $45^\circ$ slope will have a value of ${\cal I} \sim 1$.  For a near-zero value this corresponds to an incremental flow of the granular material.  It is not apparent what inertia number is relevant for regolith flows on asteroids, although different models for the migration of regolith may have values of this number at extreme limits.  For example, seismic shaking induced by impacts may yield larger values of ${\cal I}$ as the available energy is present in greater intensity and released rapidly.  Conversely, regolith motion by thermal creep may exist in a quasi-static flow regime with ${\cal I} \sim 0$.  Additional research is needed to appropriately identify and model the relevant flow regime for regolith.

The effect of the inertia number on cohesive materials has been studied numerically in \cite{rognon} and experimentally in \cite{meriaux}.  In the experimental results Meriaux created columns of cohesive material and then caused them to collapse suddenly (by removal of a supporting wall) or in a quasi-static fashion (by slowing moving a barrier wall).  The observable outcome, besides observations on the granular material morphology, were the final height of the column and the final runout length of the column.  Despite the dynamical differences between the collapse and quasi-static falls of the columns, they observed relatively consistent power law behavior between the final height and runout lengths as a function of the aspect ratio of the columns.  The implication being that, for a cohesive granular material, the inertia number is not a crucial parameter for describing the resulting flow morphology, however \cite{meriaux} notes that this is not the case for non-cohesive flows (such as dry sand).  

These same conclusions are supported by the numerical analysis presented in \cite{rognon}.  In that paper the flow dynamics and statistics were studied for a numerically evolved granular system as a function of cohesive bond number and inertia number.  They found that as bond number increased, the dynamics of the granular flow material were less sensitive to the inertia number of the flow.  The implication of these results is that, although unknown, the inertia number for the flow of regolith on an asteroid surface may not be a crucial parameter if regolith has the larger bond numbers our analysis suggests.  

\subsection{Fractures in Cohesive Materials}

One of the interesting outcomes of the experiments performed by Meriaux were the observations of stress cracks and fractures for both catastrophic and quasi-static collapse of columns of cohesive powder.  The ability of cohesive materials to mimic fractures in coherent materials has been pointed out by \cite{asphaug_LPSC} as another interpretation of the structure seen across the surface of Eros \cite{debra}.  It is instructive to scale the mechanics of stress fractures in cohesive granular materials to the asteroid environment.  In \cite{meriaux} the basic theory of stress fractures in granular materials is reviewed in a form appropriate for our use, so we rely on that paper in the following discussion.

The main parameter in determining conditions for stress fractures in granular materials is the characteristic depth $d_c$:
\begin{eqnarray}
	d_c & = & \frac{ 2 c \cos\phi}{\rho_g g_A (1-\sin\phi)}
\end{eqnarray}
where $\phi$ is the friction angle of the granular material, $\rho_g$ is the grain density, $g_A$ is the ambient acceleration, and $c$ is the cohesion of the material.  The length $d_c$ is the depth at which a granular material can undergo a stress fracture due to tension, with the plane of failure being approximately equal to the angle $\phi$ in the interior of the material.  Thus, a column of material with height on the order of $d_c$ should remain competent while a column higher than $d_c$ may begin to form cracks at this depth, which can subsequently propagate, causing collapse of the column or surface.  In general, the maximum height of a vertical slope is estimated to be twice this value \cite{meriaux}.

\begin{figure}[h!]
\centering
\includegraphics[scale=0.75]{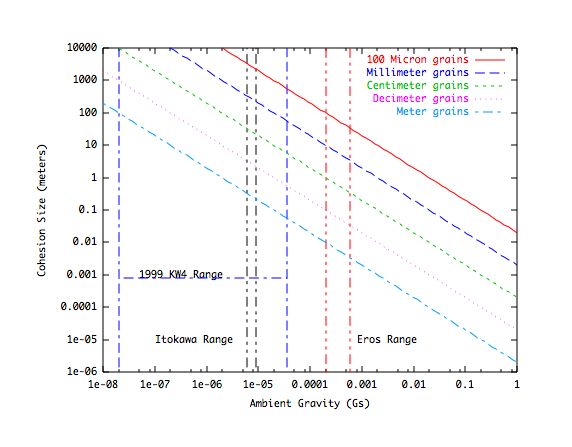}
\caption{Cohesion scale as a function of ambient gravity and particle size for lunar regolith of cleanliness $S=1$.}
\label{fig:d_c}
\end{figure}

For our purposes, we will scale the above to our previously developed force laws.  We model the cohesion with the correction incorporated by \cite{perko}
\begin{eqnarray}
	c & = & F_c \eta \\
	\eta & \sim & \frac{0.45}{r^2} \\
	F_C & \sim & 0.036 r
\end{eqnarray}
and note that the inclusion of the $\eta$ term accounts for an overall weakening of cohesive forces as a function of increasing grain size.
Using $\phi = 45^\circ$ to provide a definite estimate we find the cohesion scale length to be
\begin{eqnarray}
	d_c & \sim & 2 \times 10^{-6} \frac{1}{r G_A}
\end{eqnarray}
where $G_A$ is the ambient gravitational acceleration measured in Earth G's.  Thus, the value of $d_c$ depends on the ambient gravity and on the constituent particle size of the regolith.  In Fig.\ \ref{fig:d_c} we show the cohesion scale as a function of ambient gravity and particle size.  Also indicated on the plot are the ambient accelerations for Itokawa, Eros and 1999 KW4 Alpha.  We note that the characteristic cohesion depth as a function of particle distributions may differ from these simple extrapolations, as larger grains can have their cohesive forces weakened by smaller grains adhering to their surfaces.  These corrections are not implemented in our current analysis but can be represented by the cleanliness ratio, as mentioned previously.

For Eros we see that millimeter grains should be able to sustain structures on the order of tens of meters, and that 100 micron particles could sustain structures up to 100 meters in size.  At Itokawa, these scales increase to tens of meters for centimeter sized grains up to global structures for millimeter and smaller grains.  For the equator of KW4 we note that even meter-sized bodies can form columns tens of grains across, and could provide a different interpretation of the equatorial regions of that body.  We note that even though the size scale of these structures may be large, they will still be susceptible to episodic perturbations, such as seismic shaking, which can greatly increase the local effective ambient acceleration and cause collapse or reconsolidation.  Additionally, the upper limits for these particle sizes are idealistic as they do not incorporate the weakening effect that smaller particles will have on cohesion between large bodies, which can decrease the grain radii by a factor of 10.  

We can use these observations to motivate a reevaluation of Eros.  From the above scaling, we see that structures on the order of the grooves seen on Eros can be easily sustained and created by regolith and may be an expression of granular mechanics instead of internal structure.  Experiments show that the creation of grooves and crevices in collapsing and quasi-static flows of cohesive powders are ubiquitous and expected for sufficiently high bond numbers \cite{meriaux}.  Thus, the mechanics and dynamics of cohesive regolith flows may provide an alternate explanation for the ubiquitous groove structures seen on small bodies such as Eros \cite{procktor} and Phobos \cite{phobos_grooves}.  Instead of modeling regolith as cohesionless granular material that will flow, sand-like, into open fissures beneath the surface, we can view regolith as similar to a cohesive powder which, when subject to disturbances, can form local fractures and other features with scales potentially on the order of 100's of meters (depending on the regolith grain size) and which express intrinsic properties of the regolith itself and not necessarily deeper properties of the asteroid structure.

\subsection{The Structure of Flowing Materials}

The numerical studies of Rognon \cite{rognon} characterize the mechanics and dynamics of cohesive grains when they undergo dynamic flows.  In that study they focus on characterizing the rheological behavior of cohesive flows and how they change with bond number and inertia number.  Based on their results they make several observations on the macroscopic properties of granular flows down inclined planes.  Of specific interest for us is their conclusion that flowing cohesive granular materials will organize themselves into larger conglomerates that are then able to flow relative to each other.  The principle is simple, and can be related to the analysis of Castellanos \cite{castellanos}.  As a conglomerate grows larger its total mass increases, however the fundamental cohesive forces between it and neighboring particles are still limited by the grain size of the individual contacts.  Thus, the effective bond number of a conglomerate decreases as it grows in size, meaning that it's flow dynamics become less dominated by cohesion.  For individual grains, this is similar to their surfaces being coated by adhering gas or water vapor molecules, effectively increasing the distance between neighboring grains and decreasing their bond number.  For conglomerates the behavior should be different, however, as individual grains can be easily transported between conglomerates based on specific geometric conditions that they are subject to.  Rognon argues that this should lead to the creation of two porosity scales, one that exists within the conglomerates and one that exists between conglomerates.  While they provide some statistical results for the distribution of voids within flowing granular materials, the specific mechanics of such bi-porosity distributions has not been studied in detail in their work.  

The macroscopic implications of their work points to a specific flow morphology within cohesive granular materials.  Specifically, that cohesive granular materials will flow as larger conglomerates with a thickness characteristic of their cohesiveness on top of an under-layer of loose material, somewhat analogous to an avalanche.  Thus, one would expect cohesive regolith flows on asteroids to appear as portions of material moving as a solid and sliding to a lower region.  There are specific observations of such regolith morphology on Eros.  In \cite{veverka_regolith} a detailed analysis of bright albedo markings on the interiors of craters on Eros is given.  They conclude that these markings are due to regolith transport, with patches of regolith moving downslope due to one of several potential effects, including seismic shaking, thermal creep and electrostatic effects.  Although the downslope motion of regolith is observed in many craters at a variety of surface slopes, they are enhanced at larger slopes, greater than 25$^\circ$ in general \cite{veverka_regolith}.  The authors estimate the thickness of these flows to be less than a meter, although precise constraints are not available.  If we interpret this in terms of the characteristic depth of fractures in cohesive materials we see that this would correspond to regolith grains on the order of centimeters or less.  It is clear that the observed morphology of flows on Eros are consistent with the flow of cohesive grains, as these tend to fail and flow in surface layers sliding over a substrate that may be composed of the same material.

It is interesting to note that these same albedo markings are not seen on the asteroid Itokawa.  However, we note that the regions of finer grained regolith on that body are consistently correlated with low slope regions, potentially implying that the epoch of flow of finer materials on Itokawa has already passed for its current configuration \cite{miyamoto}, or imply that Itokawa has been subject to global shaking in the past \cite{asphaug_MAPS}.  

\subsection{Dilation and compaction of material}

Dilation is defined as the percent growth of the volume of a given granular pile.  Thus a dilation of 10\% implies an increase of volume of 10\%.  Define porosity of a granular pile as
\begin{eqnarray}
	p & = & \frac{V - V_g}{V}
\end{eqnarray}
where $V_g$ is the grain volume and $V$ is the total volume.  A fractional dilation of volume, characterized as $f = \Delta V/V$, leads to a growth in porosity of
\begin{eqnarray}
	\Delta p & = & \frac{f (1+p)}{1+f}
\end{eqnarray}
and a fractional growth in porosity of $f(1+p) / p(1+f)$.  Thus, for our 10\% dilation example ($f=0.1$) and a starting porosity of 30\% ($p=0.3$), the granular porosity would increase to 0.42, representing a 44\% increase.  Thus, for porous materials dilation leads directly to growth in porosity.  
We note that in many experiments it is not possible or easy to accurately measure the porosity of a granular aggregate, due to difficulties in measuring the total grain volume.  However, it is simple to measure dilation, as these are just measured changes in bulk volume.  Alternate definitions of porosity are  the ``solid fraction'' $\eta = 1/(1+p)$, which measures how much of an aggregates volume consists of grain volume.  In general, granular materials have a limit on their packing efficiency, which for equal sized spheres approaches a porosity of 26\% for regular packing and 37\% for irregularly packed bodies.  For aggregates composed of a distribution of sizes, minimum porosities decrease in general, as it is possible to fill interstitial gaps with smaller grains.

A fundamental property of cohesive materials is that they undergo dilation when they flow.  The experimental results of Alexander \cite{alexander} show dilation of over 20\% for avalanches of their highest cohesion material.  Similarly, Meriaux \cite{meriaux} measured dilation of up to 24\% for the quasi-static collapse of their tallest column.  In both of these measurements, however, we note that entrapped air may have contributed to overall dilation \cite{meriaux}, indicating the importance of carrying out future experiments in a vacuum chamber.

More specific results are available from numerical simulations, as it is possible to precisely compute the grain volume, initial volume and expanded volume.  Alexander is able to reproduce his observed measurements of dilation with cohesive particles with bond numbers of 45 to 90.  In their numerical experiments at bond numbers of 120 their avalanches were of the same scale as their test chamber, and hence they limited the simulated bond numbers to less than this.  

The numerical computations by Rognon et al.\ \cite{rognon} provide a much more exhaustive set of flow simulations for cohesive powders as a function of flow speed (inertia number, defined earlier) and bond number (up to 80).  They were able to precisely track all the particles in their simulations and hence provide detailed statistics.  
For flowing material at a range of speeds they find similar dilation, indicating that the amount of dilation is relatively independent of the inertia number.  In changing bond numbers from 0 to 80 they find a dilation in their flow of 25\%.  Also significant, they find heterogeneity in the distribution of pore sizes that lead to this dilation.  The standard deviation in local dilatancy, defined over a characteristic volume within the flow, ranges from 6\% for 0 bond numbers up to 14\% for bond numbers of 80, indicating that not only are there more pores distributed within the material, but that the local variation in concentration varies much more strongly in a cohesive material.  Commensurate with this, the characteristic size of pores within a cohesive distribution increases to over three times the nominal particle size with the majority of the void space being accounted for by large pores (relative to the grain size).  Hand-in-hand with the distribution of pores is grain clumping, which forms aggregates of increasing size with increased cohesion.  These effects are supported by the increased ability of cohesive grains to maintain themselves in a group.

The results of dilation in flows can also be reversed by addition of seismic energy or ``tapping.''  This will cause a dilated distribution to shrink, closing up the pores opened during a previous period of flow.  In terms of ``packing fraction'' $\eta$, the derived law for compaction can be stated as:
\begin{eqnarray}
	\eta(n) & = & \eta_{\infty} - \frac{\eta_{\infty} - \eta_o}{1 + B \ln{\left(1 + \frac{n}{\tau}\right)}}
\end{eqnarray}
where $B$ and $\tau$ are empirically derived quantities, $n$ is the number of taps, $\eta_o$ is the initial packing fraction and $\eta_{\infty}$ is the limiting packing fraction.  Such relationships have been verified for cohesive powders as well as for non-cohesive materials \cite{vandewalle}.  Indications are that cohesive powders can actually experience larger relative compactions, due perhaps to the inter-particle cohesive forces and to the initially larger dilation amounts that they can obtain.  This seemingly reversible process can thus also cause compaction of regolith and, depending on the environment in which the regolith is placed, could yield a larger bulk density.

\section{Discussion}

This paper has a few specific purposes.  First is to establish and compare the different forces that are relevant for regolith on the surfaces of small asteroids.  From this comparison we identify cohesion as a potentially important physical force for these systems.  Second is to reinterpret the existing literature on cohesive granular mechanics in terms of granular mechanics phenomenon on asteroid surfaces.  This is not easily done, given the large scale differences between terrestrial labs and the asteroid environment, and that these experiments have not been designed to recreate certain crucial elements of the asteroid environment such as vacuum and lack of trace gases and water vapor.  Still, the comparison seems to show some merit and should hopefully motivate new studies of cohesive powders that may be related more directly to the asteroid environment.  Finally, in the following we take a slightly larger view and reconsider a few basic ideas and tenets regarding asteroids and their interpretation, and explore the implications of viewing these systems from a cohesive granular mechanics point of view.  Specifically we discuss the possible implications for interpreting asteroid surface imagery, porosity distribution within asteroids, and finally reconsider the terminal evolution of small asteroids subject to the YORP effect.

\subsection{Implications for interpretations of asteroid surface imagery}

Previous views of asteroid surfaces have been limited in their spatial resolutions to centimeters at best, and then only over extremely limited regions at a fixed, relatively low phase angle \cite{veverka_landing, yano}.  Similarly, the surfaces of Phobos and Deimos, the other small asteroid-like bodies that have detailed shape imagery, have even lower spatial resolutions \cite{thomas_phobos_deimos}.  Despite this limitation, there is ample evidence for finer regolith grains at the sub-centimeter and smaller level on all of these bodies.  In previous literature, the surfaces of these bodies has usually been interpreted using the terminology and physics of terrestrial geology.  For example, on Eros the ubiquitous lineaments and other surface structures have been interpreted as expressing sub-surface strength features \cite{debra, procktor} while on Itokawa the surface has been analyzed in terms of landslide phenomenon as found on Earth \cite{miyamoto}.  
If, instead, we apply the results described in this paper, essentially following the suggestions in \cite{asphaug_LPSC}, and interpret these surfaces in light of cohesive forces and their effects, we may arrive at an alternate array of conclusions.  

For understanding the visible structures on Itokawa, we realize that the seemingly dominant grain size in the Muses-sea region (on the order of centimeters in size) may actually be agglomerates of smaller materials which have formed into this characteristic size during their flow down to the potential lows of the system.  This scenario is consistent with the flow dynamics of cohesive powders, as detailed in \cite{rognon}, in which they preferentially clump into larger aggregates which can then travel more freely, mimicking larger grains as they undergo transport.  The ability of these aggregate structures to maintain their shape over long time periods in the space environment is not known nor has it been studied.  Laboratory tests with cohesive powders could shed light on this potential phenomenon, by studying flows of cohesive powders in vacuum conditions, and subsequently studying the mechanics of these systems subjected to repeated tapping that would mimic seismic shaking.  Our statements do not preclude the presence of larger, coherent particles that are not held together with cohesive forces, but does indicate that based on our scaling arguments one cannot exclude the possibility that some of these structures may also be constructed out of conglomerates.

Next we consider Eros.  Based on our scaling laws we note that the large scale structures on Eros have the appropriate size to also be interpreted as expressions of regolith strength and fracture due to cohesion, instead of loose material expressing the structure of bedrock beneath its surface.  
Again, the literature on geophysics of cohesive powders is relatively non-existent, however if we consider the basic mechanics of cohesive powders we can directly propose other possible mechanisms for the formation of surface structures on regolith covered asteroids such as Eros.
Specifically, we propose that the formation of surface lineaments could be due to stresses induced by dilation of material either beneath the surface or of the regolith itself as an outcome of being subjected to transmitted seismic waves.
Given the universal nature of dilation with flow, it would be of special interest to better understand the effect of seismic waves on cohesive grains.  By definition, the S-waves that occur following a seismic event represent macroscopic motion of individual grains and hence could lead to a dilation of the material with subsequent changes in the surface stress field, which could lead to fracture and other surface expressions.  Conversely, P-waves generally represent the transmission of pressure without motion, and hence could represent ``tapping'' phenomenon which is known to be able to reduce porosity in a cohesive material \cite{vandewalle}.  

Also affected by these apparent cohesive forces are the proposed mechanisms for dust levitation and migration on the surfaces of asteroids \cite{lee, colwell}.  By directly comparing cohesive forces to the enhanced electrostatic forces that may arise on occasion at an asteroid's terminator we see that the cohesive forces generally dominate until one arrives at the few millimeter size-scale or larger.  Previous suppositions have assumed that dust particles on the order of tens to hundreds of microns were the primary components of levitated particles.  If this remains true, we require some other mechanism for breaking the cohesive bonds between such small grains, such as micro-meteoroid impacts, with the predicted amount of levitated dust being significantly reduced by the enhanced ability of the materials to adhere to each other and the highly localized conditions that can generate such strong electric fields.  This leads to a view of asteroid surfaces where they remain dominated by smaller-scale dust particles that adhere to each other to form larger conglomerates.  This model would be consistent with the measurements reported by Masiero \cite{masiero} which found evidence of a uniform structure for asteroid surfaces at the small size scale, and did not detect any evidence for the depletion of smaller particles.

These hypotheses can be probed at three different levels.  First, and most basic, they can be tested by sending a space science mission to the surface of a small body in order to carry out high spatial resolution imaging, preferably at the sub-millimeter size scale, in order to observe the morphology of the smallest components on the surface.  Associated with such an exploration should also be tests of the mechanical strength of the surface components, which could either be carried out with a portable lab or through observations of the surface probe interactions with the asteroid surface.  

The other two approaches can be carried out on Earth.  First would be laboratory experiments using cohesive powders with bond numbers and particle size distributions chosen to mimic models of asteroid regolith distributions.  In contrast with current studies, these should be specialized to better mimic the asteroid surface, such as the use of vacuum chambers, high temperatures to clear water vapor, and under appropriate illumination and electrostatic charging environments.  Specific items for study would be the global reshaping of cohesive powders due to intermittent shaking, the seismic transmission of waves through cohesive powders and avalanche morphology and mechanics in cohesive powders.  Some researchers have initiated tests of granular material in the appropriate environments \cite{blum}, and these experiments would serve as an excellent starting point for additional research.

Last is the numerical simulation of granular mechanics using appropriate models for size distribution and environment.  These are, perhaps, the most easily accessed.  An appropriate starting point for such investigations is found in \cite{sanchez} which discusses the application of granular mechanics techniques to the asteroid environment.  The questions of interest are the same as above, however in a computational environment it is often possible to gain deeper insight into the statistics of these processes, at the cost of realism.

\subsection{Implications for interpretations of asteroid Micro and Macro Porosity}

A second and significant implication of our study relates to the distribution of porosity within asteroids.  Since the first precise asteroid mass determination of Mathilde showed that body to have significant porosity, potentially greater than 50\% \cite{mathilde}, it has been firmly established that asteroids can exhibit high degrees of porosity.  What is not clear is how that porosity is distributed within an asteroid, as micro-porosity or as a few large voids in the interiors of an asteroid.  There have been a number of models proposed to describe how this porosity may arise and persist, however it is difficult to test any of these models and their implications \cite{B&C_AIII}.  

One motivating result from our current analysis is a clear link between the expected physics of a granular media in the asteroid environment and readily observed dilation and high levels of porosity that can be easily created in cohesive granular materials that undergo flow -- either catastrophic or quasi-static \cite{meriaux}.  The reversibility of this process is also interesting, as it can lead to bodies of similar composition having a range of porosities as a function of the processes which occur on them.  

An additional element that is present on asteroids is their peculiar geometries.  It is known that at modest spin rates, loose materials on an asteroid will preferentially flow to the polar regions, while at high spin rates they will migrate to the equator \cite{guibout}.  Such a contrast is explicitly seen on Itokawa and 1999 KW4 Alpha.  The phenomenon is due to the change of the surface geopotential lows.  We can note that the geometry which the regolith encounters at the polar regions is markedly different to that found at the equator, however.  

Material that flows to a polar region, or for a more specific example to the Muses Sea region of Itokawa, are entering a confined region \cite{miyamoto}.  Even if the flows undergo dilation, they will be compressing previous flows into the same geometric region and hence may become compacted.  For Itokawa this scenario is consistent with the relatively compacted surface reported in \cite{yano} and in the non-uniform density inferred in \cite{itokawa_mdist}.  The secondary of the 1999 KW4 system also has a relatively higher density as compared to the primary, which could similarly result from its slow rotation (meaning that the polar regions are the geopotential low) and its continuous shaking \cite{KW4}.

The situation is much different for material that flows to the equatorial region of a fast rotator.  In this situation, as material flows to the equator, it can achieve a lower position in the geopotential well by increasing its distance from the body.  Due to this, material is free to expand into an unconfined region, which may enable flow-induced dilation to remain present in the materials as they are not subject to compression.  Expansion is only limited by the synchronous orbit locations above the surface, as passage through that point will place the grain into orbit.  
We note that the synchronous orbit locations on 1999 KW4 Alpha are on the order of meters above the equator and could be at the surface at at least two points within the model uncertainty.  This situation is also consistent with the low density of Alpha relative to Beta, and the consequent high porosity varying between 40\% and 66\%.  This corresponds either to a uniformly under-dense body or a body with usual porosity and a region of very high porosity.  This is consistent with the equatorial bulge on this body which also has extremely low ambient gravity (less than a tenth of a micro-gravity), implying that cohesive effects are important for bodies on the order of tens of centimeters in size.  Any disturbances that may travel through this regime will displace particles towards a lower gravitational environment that has no hard constraint, unlike the situation in the seas of Itokawa.  Thus, we can tender a hypothesis that the equatorial region of 1999 KW4 Alpha is a low density region which has undergone substantial dilation.
Finally, we hypothesize that a low porosity for an asteroid may correspond to a past period of high rotation, where such a reversal in the geopotential would have occurred.

\subsection{A new model for the terminal evolution of small asteroids}

Finally, we consider the implication of our findings for the evolution of small asteroids as they are spun-up by the YORP effect.  While visual imagery of asteroid surfaces shows an abundance of boulders at size scales of meters to tens of meters at Itokawa, there is no direct evidence for monolithic components on that body at size scales of 100 meters, which is still below the cut-off for fast spinning asteroid sizes.  Similarly, while Eros has a number of clearly defined blocks on its surface approaching 100 meters in size, the vast amount of material contained (at the surface of that body at least) lies in the much finer regolith that blankets the body.  We note that Eros is also an exceptional body, as its YORP time scale is very long, due to its large size, and hence it is likely that the YORP effect has not had a dramatic influence on the evolution of this body.

We focus ourselves on asteroids that are subject to the YORP effect, in general 5 km radius bodies and smaller in the NEO population and perhaps up to a few tens of km in the main belt.
Based on the images from Itokawa and the observed spin-fission limits, we presume that larger asteroids consist of distributions of boulders of all shapes and sizes.  When subject to YORP they spin faster and, following from basic celestial mechanics principles \cite{Icarus_fission,PSS_CD}, can shed their largest components into orbit when their spins become rapid enough (which can be significantly less than the traditional spin-fission rotation period limit of $\sim 2.3$ hours if the body has a strong binarity to its shape).  Loss of these components can change the YORP torques and either reinforce the loss process or provide a hiatus when the body undergoes a spin down and spin up YORP cycle.  There are limits on the size of a component which can be directly shed, as components with a mass fraction larger than $\sim$0.17 will have a negative total energy and must undergo further splitting to be ejected on a short time scale. These can form binaries or reimpact to become contact binaries.  Boulders or aggregates with a mass fraction smaller than 0.17 will be subject to relatively rapid ejection from the system \cite{scheeres_F2BP_planar}.

Repetition of this process can gradually remove the largest competent boulders or agglomerates from an asteroid, while preferentially retaining the smaller, and hence more cohesive, grains.  From our previous discussions, we note that centimeter-sized grains in proximity to each other can provide sufficient cohesive force to withstand a few-minute period rotation rate of a 100 meter asteroid.  For a 10 meter body similar grain sizes could withstand a rotation period of less than a minute.  Combining these two effects -- the preferential loss of larger components on a body spun to high rotation rates and the preferential cohesion between smaller regolith grain sizes -- we can envision that small NEOs undergo a fractionation process that liberates larger boulders and retains finer regolith on the remaining largest component.  
In particular, we refer again to Fig.\ \ref{fig:antigravity} and note that positive accelerations at the surface of a 100 meter asteroid are only 0.1 milli-Gs for a half-hour rotation period and 1 milli-G for a six minute rotation period, and a 10 meter asteroid spinning with a period of less than a minute has a 1 milli-G positive acceleration.  The cohesion force balances this positive acceleration for grain radii of 1 centimeter or less.  
This again reinforces the fact that rapidly spinning asteroids and rubble-pile asteroids are not mutually exclusive, as has been asserted in a number of previous papers by Holsapple \cite{holsappleA, holsappleB,holsapple_smallfast}.

As this process continues, the ever-smaller components are more strongly affected by YORP and should undergo cycles with increasing frequency, and hence be susceptible to the loss of additional components at an increasing pace.  Even if the interior of an asteroid such as Itokawa contain some larger monolithic components, the fractionation of these bodies should eventually expose these interior objects, as failure should preferentially occur along larger grains due to their decreased cohesion.  After this, they will become the next component of the asteroid to be shed when the spin rate increases to the appropriate rate.  The details of this process are likely more complex, as the relative size of the two components strongly controls the subsequent evolution of these systems \cite{scheeres_F2BP_planar} and can even lead to systems that remain ``stuck'' in a contact binary cycle for long periods of time \cite{Icarus_fission}.

Such rapidly spinning aggregates would also be susceptible to fracture, however, as a micro-meteorite impact could break cohesive bonds between conglomerates within these bodies.  Given our knowledge of the physics of cohesion, such a fracture would not cause the aggregate to uniformly disrupt, but would cause it to fail along naturally occurring stress fractures within the aggregate, as occurs when cohesive powders fail \cite{meriaux}.  The mechanical outcome of such a fracture would keep the components rotating at their same rate initially and would decrease the accelerations the grains are subject to due to the decreased body size.  However the large changes in mass distribution would cause the components to immediately enter a tumbling rotation state.  It is significant to note that there is evidence for some small, rapidly rotating bodies to be in tumbling rotation states \cite{pravec_tumbling} -- such fracturing consistent with cohesive materials could provide an explanation for this.  

One prediction of this model is that small asteroids may be formed both from monolithic boulders as well as from cohesive gravels of small enough size.  The properties of these different morphology types, such as thermal inertia or polarization, should be investigated and the relative abundance of monoliths or cohesive gravels among the small body population would mimic the size distribution of fractured asteroids.  
Finally, this model of spinning cohesive gravels is consistent with the Holsapple results, but may provide a clearer physical mechanism for how the small amounts of strength required by the Holsapple models manifest themselves.  

\section{Conclusions}

We study the relative effect of gravitational and non-gravitational forces in the asteroid environment and find that cohesive forces may play an important role for these bodies.  We review some of the experimental and computational research literature on the mechanics of cohesive powders and find interpretations that can shed light on possible physical phenomenon at asteroids.  We consider implications of this research and point out future experiments and hypotheses concerning the importance of cohesive forces that can be tested.  Finally, we propose reinterpretations of asteroid observations and populations in light of these cohesive forces.  This process leads to significantly different conclusions for the geophysical properties of asteroid surfaces, interiors and of small rapidly rotating asteroids.


\begin{thebibliography}{199}

\bibitem{alexander}
A.W.\ Alexander et al.  2006.  ``Avalanching flow of cohesive powders,'' {\it Powder Technology} 164: 13-21.

\bibitem{asphaug_LPSC}
E.I. Asphaug.  2009.  ``Shattered dirt:  Surface fracture of granular asteroids,'' 40th Lunar and Planetary Science Conference.  Abstract \# 1438.

\bibitem{asphaug_MAPS}
E. Asphaug. 2008.
``Critical crater diameter and asteroid impact seismology,''
{\it Meteoritics \& Planetary Science} 43(6): 1075Ð1084.

\bibitem{blum}
P.G. Hofmeister, J. Blum, D. Heisselmann. 2009.
``The Flow Of Granular Matter Under Reduced-Gravity Conditions,''
arXiv:0905.0330v2 [astro-ph.EP] 12 May 2009.

\bibitem{B&C_AIII}
D.T. Britt, D. Yeomans, K. Housen and G. Consolmagno. 2002.
``Asteroid Density, Porosity and Structure,'' in \underline{Asteroids III}, (W.F. Bottke, Jr., A. Cellino, P. Paolicchi and R.P. Binzel Eds.),  University of Arizona Press, Tucson.

\bibitem{debra}
D.L. Buczkowski, O.S. Barnouin-Jha, and L.M. Prockter.  2008.  ``433 Eros lineaments: Global mapping and analysis,'' {\it Icarus} 193: 39-52.

\bibitem{burns}
J.A. Burns, P.L. Lamy, and S. Soter. 1979.
``Radiation forces on small particles in the solar system,'' {\it Icarus} 40(1): 1 - 48.

\bibitem{castellanos}
A. Castellanos.  2005.
``The relationship between attractive interparticle forces
and bulk behaviour in dry and uncharged fine powders,'' {\it Advances in Physics} 54(4): 263-376.

\bibitem{criswell}
D.R. Criswell and B.R. De. 1977. 
``Intense Localized Photoelectric Charging in the Lunar Sunset Terminator Region, 2. Supercharging at the Progression of Sunset,''
{\it Journal of Geophysical Research} 82(7): 1005-1007.

\bibitem{colwell}
J.E. Colwell, A.A.S. Gulbis, M. Hor\'anyi, and S. Robertson. 2005.
``Dust transport in photoelectron layers and the formation of dust ponds on Eros,''
{\it Icarus} 175(1): 159-169.

\bibitem{lunar_review}
J. E. Colwell, S. Batiste, M. Hor\'anyi, S. Robertson, and S. Sture. 2007.
``Lunar Surface:  Dust Dynamics and Regolith Dynamics,'' 
{\it Reviews of Geophysics} 45, RG2006.

\bibitem{dankowicz}
H. Dankowicz.  ``Some special orbits in the two-body problem with radiation pressure,'' {\it Celestial Mechanics and Dynamical Astronomy} 58:  353-370, 1994.

\bibitem{demura}
H.\ Demura, S.\ Kobayashi, E./ Nemoto, N.\ Matsumoto, M.\ Furuya, A.\ Yukishita, N.\ Muranaka, H.\ Morita, K.\ Shirakawa, M.\ Maruya, H.\ Ohyama, M.\ Uo, T.\ Kubota, T.\ Hashimoto, J.\ Kawaguchi, A.\ Fujiwara, J.\ Saito, S.\ Sasaki, H.\ Miyamoto, and N.\ Hirata.
``Pole and Global Shape of 25143 Itokawa,'' {\it Science} 312:  1347-1349.

\bibitem{derjaguin}
B.V. Derjaguin, V.M. Muller and Y.P. Toporov.  1975.
``Effect of Contact Deformations on the Adhesion of Particles,'' {\it Journal of Colloid and Interface Science} 53(2): 314-326.

\bibitem{fujiwara}
 A. Fujiwara, J. Kawaguchi, D. K. Yeomans, M. Abe, T. Mukai, T. Okada, J. Saito, H. Yano, M. Yoshikawa, {D. J. Scheeres},  O. Barnouin-Jha, A. F. Cheng, H. Demura, R. W. Gaskell, N. Hirata, H. Ikeda, T. Kominato, H. Miyamoto, A. M. Nakamura, R. Nakamura, S. Sasaki, and K. Uesugi. 2006.
``The Rubble-Pile Asteroid Itokawa as Observed by Hayabusa,'' {\it Science} 312:  1330-1334.

\bibitem{guibout}
{ V. Guibout} and {D.J. Scheeres}.  2003. ``Stability of Surface Motion on a Rotating Ellipsoid,'' {\it
Celestial Mechanics \& Dynamical Astronomy} 87: 263-290.

\bibitem{heim}
L.-O. Heim, J. Blum, M. Preuss, and H.-J. Butt.  1999.
``Adhesion and Friction Forces between Spherical Micrometer-Sized Particles,''
{\it Physical Review Letters} 83(16): 3328-3331.

\bibitem{hermann}
H. J. Herrmann and S. Luding.  1998.  ``Modeling granular media on the computer,'' {\it Continuum Mech. Thermodyn.} 10: 189-231.

\bibitem{holsappleA} 
K.A. Holsapple.   2001.  ``Equilibrium Configurations of Solid Cohesionless Bodies,'' {\it Icarus} 154:  432-448.

\bibitem{holsappleB} 
K.A. Holsapple.   2004.  ``Equilibrium figures of spinning bodies with self-gravity,'' {\it Icarus} 172:  272-303.

\bibitem{holsapple_smallfast}
K.A. Holsapple. 2007. ``Spin limits of Solar System bodies:
From the small fast-rotators to 2003 EL61,'' {\it Icarus} 187: 500 - 509.

\bibitem{holsapple_deform}
K.A. Holsapple. 2009. ``On YORP-induced spin deformations of asteroids,'' {\it Icarus} doi:10.1016/j.icarus.2009.08.014

\bibitem{hughes}
A.L.H. Hughes, J.E. Colwell, and A. Ware DeWolfe. 2008.
``Electrostatic dust transport on Eros: 3-D simulations of pond formation,''
{\it Icarus} 195(2): 630-648.

\bibitem{johnson}
K. L. Johnson, K. Kendall and A. D. Roberts. 1971.
``Surface Energy and the Contact of Elastic Solids,''
{\it Proceedings of the Royal Society London A} 324: 301-313.

\bibitem{ask}
A.S. Konopliv, J.K. Miller, W.M. Owen, D.K. Yeomans, J.D. Giorgini, R. Garmier and J.-P. Barriot.  2002.  ``A Global Solution for the Gravity Field, Rotation, Landmarks,
and Ephemeris of Eros,'' {\it Icarus} 160: 289-299.

\bibitem{lee}
P. Lee. 1996.
``Dust Levitation on Asteroids,''
{\it Icarus} 124(1): 181-194.

\bibitem{veverka_regolith}
A. Mantz, R. Sullivan, and J. Veverka. 2004.
``Regolith transport in craters on Eros,''
{\it Icarus} 167: 197-203.

\bibitem{marshall}
J.R. Marshall, T.B. Sauke, and J.N. Cuzzi.  2005.
``Microgravity studies of aggregation in particulate clouds,''
{\it Geophysical Research Letters} 32, L11202.

\bibitem{masiero}
J. Masiero, {C. Hartzell}, and {D.J. Scheeres}.  2009.  ``The effect of the dust size distribution on asteroid polarization,'' {\it The Astronomical Journal} 138: 1557-1562.

\bibitem{maugis}
D. Maugis. 1992.
``Adhesion of Spheres: The JKR-DMT Transition Using a Dugdale Model,''
{\it Journal of Colloid and Interface Science} 150(1): 243-269.

\bibitem{meriaux}
C. M\'eriaux and T. Triantafillou.  2008.  ``Scaling the final deposits of dry cohesive granular columns after collapse
and quasi-static fall,'' {\it Physics of Fluids} 20:  033301.

\bibitem{michikami}
T. Michikami, A.M. Nakamura, N. Hirata, R.W. Gaskell, R. Nakamura, T. Honda, C. Honda, K. Hiraoka, J. Saito, H. Demura, M. Ishiguro, and H. Miyamoto. 2008.
``Size-frequency statistics of boulders on global surface of asteroid 25143 Itokawa,''
{\it Earth, Planets and Space} 60: 13-20.

\bibitem{mignard}
F. Mignard and M. H\'enon.  ``About an unsuspected integrable problem,'' {\it Celestial Mechanics and Dynamical Astronomy} 33: 239-250, 1984.  

\bibitem{miller}
J.K Miller, A.S. Konopliv, P.G. Antreasian, J.J. Bordi, S. Chesley, C.E. Helfrich, W.M. Owen, T.C. Wang, B.G.
Williams, D.K. Yeomans, and {D.J. Scheeres}.  2002.  ``Determination of shape, gravity and rotational state of
Asteroid 433 Eros,'' {\it Icarus} 155: 3-17.

\bibitem{miyamoto}
H. Miyamoto, H. Yano, {D.J. Scheeres}, S. Abe, O. Barnouin-Jha, A.F. Cheng, H. Demura, R.W. Gaskell, N. Hirata, M. Ishiguro, T. Michikami, A.M. Nakamura,  R. Nakamura, J. Saito, and S. Sasaki.
2007. ``Regolith migration and sorting on asteroid Itokawa,''
{\it Science} 316:  1011-1014. 

\bibitem{pingpong}
K. Nishimura, S. Keller, J. McElwaine, Y. Nohguchi.  1998.  ``Ping-pong ball avalanche at a ski jump,'' {\it Granular Matter} 1:  51-56.

\bibitem{ostro_radar}
S.J. Ostro, R.S. Hudson, L.A.M. Benner, J.D. Giorgini, C. Magri, J.-L. Margot, M.C. Nolan.  2002.  ``Asteroid Radar Astrometry,'' in \underline{Asteroids III}, (W.F. Bottke, Jr., A. Cellino, P. Paolicchi and R.P. Binzel Eds.),  University of Arizona Press, Tucson.

\bibitem{KW4_ostro}
S. J. Ostro, 
J.-L. Margot, 
L. A. M. Benner, 
J. D. Giorgini, 
{D. J. Scheeres}, 
{E. G. Fahnestock}, 
{S. B. Broschart}, 
{J. Bellerose}, 
M. C. Nolan, 
C. Magri, 
P. Pravec, 
P. Scheirich, 
R. Rose, 
R. F. Jurgens, 
S. Suzuki, 
E. M. DeJong. 2006. ``Radar Imaging of Binary Near-Earth Asteroid (66391) 1999 KW4,'' {\it Science} 314: 1276-1280.

\bibitem{perko}
H.A. Perko, J.D. Nelson, and W.Z. Sadeh.  2001.
``Surface Cleanliness Effect on Lunar Soil Shear Strength,''
{\it Journal of Geotechnical and Geoenvironmental Engineering} 127(4): 371-383.

\bibitem{HarrisPravec}  
P. Pravec and A.W. Harris.  2000.  ``Fast and Slow Rotation of Asteroids,'' {\it Icarus} 148:  12-20.

\bibitem{pravec_tumbling}
P. Pravec, et. al.  2005.
``Tumbling asteroids,'' {\it Icarus} 173: 108-131.

\bibitem{pravecharris2006}
P. Pravec, et al.\ 2006.  ``Photometric survey of binary near-Earth asteroids,'' {\it Icarus} 181:  63-93.

\bibitem{procktor}
L. Procktor, et al. 2002.
``Surface Expressions of Structural Features on Eros,'' {\it Icarus} 155: 75-93.

\bibitem{richardson}
D.C. Richardson, P. Elankumaran, and R.E. Sanderson.  2005.  ``Numerical experiments with rubble piles: equilibrium shapes and spins,'' {\it Icarus} 173: 349-361.

\bibitem{richardson_cohesion}
D.C. Richardson, P. Michel, K.J. Walsh, and K.W. Flynn.  2009.  ``Numerical simulations of asteroids modelled as gravitational aggregates with cohesion,'' {\it Planetary and Space Science} 57(2): 183-192.

\bibitem{jim_richardson}
J.E. Richardson, H.J. Melosh and R. Greenberg.  2004.  ``Impact-Induced Seismic Activity
on Asteroid 433 Eros: A Surface
Modification Process,'' {\it Science} 306: 1526-1529.

\bibitem{richter}
K. Richter and H.U. Keller.  ``On the stability of dust particle orbits around cometary nuclei,'' {\it Icarus} 114:  355-371, 1995.

\bibitem{robinson}
M.S. Robinson, P.C. Thomas, J. Veverka, S. Murchie and B. Carcich. 2001.
``The nature of ponded deposits on Eros,''
{\it Nature} 413: 396-400.

\bibitem{rognon}
P.G. Rognon, J.-N. L. Roux, M. Naaim, and F. Chevoir.   2008.  ``Dense flows of cohesive granular materials,'' {\it Journal of Fluid Mechanics} 596: 21-47.

\bibitem{sanchez}
P. S\'{a}nchez and {D.J. Scheeres}.  2009.  ``Granular mechanics in asteroid regolith:  Simulating and scaling the brazil nut effect,'' abstract \# 2228, presented at the 40th Lunar and Planetary Science Conference.



\bibitem{D}  
D.J. Scheeres.  1999.  ``Satellite Dynamics about Small Bodies:  Averaged Solar Radiation Pressure Effects,'' Journal of the Astronautical Sciences 47(1):  25-46.

\bibitem{scheeresAIII} 
{D.J. Scheeres}, D.D. Durda, and P.E. Geissler. 2002. The Fate of Asteroid
Ejecta, in {\it Asteroids III} (W.M. Bottke Jr., A. Cellino, P. Paolicchi, R.P. Binzel eds.),
University of Arizona Press, Tucson, pp. 527-544.

\bibitem{lpsc_SRP}
{D.J. Scheeres}.  ``Solar Radiation Pressure and Transient Flows on Asteroid Surfaces [\#1919],'' abstract presented at the
Lunar and Planetary Science XXXVI meeting, Houston, Texas, March 2005.

\bibitem{KW4} 
{D. J. Scheeres}, 
{E. G. Fahnestock}, 
S. J. Ostro, 
J.-L. Margot, 
L. A. M. Benner, 
{S. B. Broschart}, 
{J. Bellerose}, 
J. D. Giorgini, 
M. C. Nolan, 
C. Magri, 
P. Pravec, 
P. Scheirich, 
R. Rose, 
R. F. Jurgens, 
S. Suzuki, 
E. M. DeJong. 2006.   ``Dynamical Configuration of Binary Near-Earth Asteroid (66391) 1999 KW4,'' {\it Science} 314:  1280-1283.

\bibitem{Icarus_fission}
{D.J. Scheeres. }  2007.
``Rotational fission of contact binary asteroids,'' {\it Icarus} 189:  370-385.

\bibitem{itokawa_mdist}
{D.J. Scheeres} and R.W. Gaskell.  2008.  ``Effect of density inhomogeneity on YORP:  The case of Itokawa,'' {\it Icarus} 198:  125-129.

\bibitem{PSS_CD}
{D.J. Scheeres}.  2009.  ``Minimum energy asteroid reconÞgurations 
and catastrophic disruptions,'' {\it Planetary and Space Science} 57:  154-164.

\bibitem{scheeres_F2BP_planar}
{D.J. Scheeres}. 2009.  
``Stability of the Planar Full 2-Body Problem,'' {\it Celestial Mechanics and Dynamical Astronomy} 104: 103-128.

\bibitem{schwedes}
J. Schwedes.  1975.
``Shearing Behaviour of Slightly Compressed Cohesive Granular Materials,''
{\it Powder Technology} 11: 59-67.

\bibitem{sharma}
I. Sharma, J.T. Jenkins, J.A. Burns.  2009.
``Dynamical passage to approximate equilibrium shapes for spinning, gravitating rubble asteroids,'' {\it Icarus} 200: 304-322.

\bibitem{phobos_grooves}
P. Thomas, J. Veverka, A. Bloom and T. Duxbury.  1979.
``Grooves on Phobos:  Their Distribution, Morphology and Possible Origin,''
{\it Journal of Geophysical Research} 84(B14): 8457-8477.

\bibitem{thomas_phobos_deimos}
P. Thomas. 1979.
``Surface features of Phobos and Deimos,'' {\it Icarus} 40(2): 223-243.

\bibitem{Eros_crater_thomas} 
P.C. Thomas and M.S. Robinson.  2005. ``Seismic resurfacing by a single impact on the
asteroid 433 Eros,'' {\it Nature} 436: 366-369.

\bibitem{vandewalle}
N. Vandewalle, G. Lumay, O. Gerasimov, and F. Ludewig.  2007.  ``The influence of grain shape, friction and cohesion on granular
compaction dynamics,'' {\it The European Physical Journal E} 22: 241Ð248.

\bibitem{veverka_landing}
J. Veverka et al.  2001.  ``The landing of the NEAR-Shoemaker
spacecraft on asteroid 433 Eros,'' {\it Nature} 413: 390-393.

\bibitem{wilkison}
S.L. Wilkison, M.S. Robinson, P.C. Thomas, J. Veverka, T.J. McCoy, S.L. Murchie, L.M. Prockter, D.K. Yeomans. 2002. ``An estimate of Eros's porosity and implications
for internal structure,'' {\it Icarus} 155: 94-103.

\bibitem{yano}
 H. Yano, T. Kubota, H. Miyamoto, T. Okada, {D. J. Scheeres},  Y. Takagi, K. Yoshida, M. Abe, S. Abe, O. Barnouin-Jha, A. Fujiwara, S. Hasegawa, T. Hashimoto, M. Ishiguro, M. Kato, J. Kawaguchi, T. Mukai, J. Saito, S. Sasaki, and M. Yoshikawa. 2006.
``Touch-down of the Hayabusa spacecraft at the Muses Sea on Itokawa,'' {\it Science} 312:  1350-1353. 

\bibitem{mathilde}
D.K. Yeomans, J.-P. Barriot, D.W. Dunham, R.W. Farquhar, C.L. Helfrich,
A.S. Konopliv, J.V. McAdams, J.K. Miller, {D.J. Scheeres}, S.P. Synnott,
W.M. Owen, and B.G. Williams.  1997.  ``The NEAR Spacecraft's Flyby of Asteroid 253
Mathilde,'' {\it Science} 278:2106--9.









\end{thebibliography}
\end{document}